\documentclass[useAMS,usenatbib]{mnras}
\usepackage[latin1]{inputenc}
\usepackage[english]{babel}
\usepackage{graphicx}
\usepackage{amsmath}
\usepackage{amsfonts}
\title[Cosmic Acceleration in an Extended Brans-Dicke-Higgs Theory]{Cosmic Acceleration in an Extended Brans-Dicke-Higgs Theory}
\author[Soumya Chakrabarti]{Soumya Chakrabarti\thanks{E-mail : soumya.chakrabarti@saha.ac.in}\\
Theory Division\\
Saha Institute of Nuclear Physics\\
Kolkata 700064\\
India}
\date{Accepted XXX. Received YYY; in original form ZZZ}
\pubyear{2020}
\begin{document}
\maketitle
\begin{abstract}
We consider an extended scalar-tensor theory of gravity where the action has two interacting scalar fields, a Brans-Dicke field which makes the effective Newtonian constant a function of coordinates and a Higgs field which has derivative and non-derivative interaction with the lagrangian. There is a non-trivial interaction between the two scalar fields which dictates the dominance of different scalar fields in different era. We investigate if this setup can describe a late-time cosmic acceleration preceded by a smooth transition from deceleration in recent past. From a cosmological reconstruction technique we find the scalar profiles as a function of redshift. We find the constraints on the model parameters from a Markov Chain Monte Carlo analysis using observational data. Evolution of an effective equation of state, matter density contrast and thermodynamic equilibrium of the universe are studied and their significance in comparison with a $\Lambda$CDM cosmology is discussed.
\end{abstract}
\section{Introduction}\label{s0}
General Theory of Relativity (GR) carries the badge of being the most celebrated theory of gravity even after a century of it's foundation. However, it has left behind some philosophical gaps, inspiring physicists to look for a generalized or `better' theory of gravity. Brans and Dicke made a novel attempt to write such a generalized theory. Their theory (BD theory) did combine Mach's Principle with gravity ensuring that any local motion is a function of large-scale matter distribution of the Universe \citep{BD}. In BD theory a scalar field, henceforth called the BD scalar, interacts non-minimally with curvature and makes the effective Newtonian constant $G$ a function of the scalar field \citep{fierz, BD}. The BD coupling parameter $\omega$ serves as a signature of the theory. The BD field equations reduce to GR equations for a large $\omega$ limit \citep{will}. However, the large $\omega$ limit also leads to severe constraints, as the theory effectively fails to work for any general stress-energy tensor in the aforementioned limit \citep{nb1, faraoni}. Regardless, the theory receives due attention as it can drive the enigmatic cosmic acceleration naturally \citep{Gut81, Rie98, Per99}. The theory shows a way out of the so-called {\it `graceful exit'} problem in models of early inflation through the idea of extended inflation \citep{mathi, la}. Moreover, the geometric scalar field in BD theory can also generate the effective negative pressure required to drive the late-time acceleration \citep{nb2, sudiptadi, mota, sensen}, avoiding the arbitrariness of putting in an exotic matter field by hand. \\

High precision astrophysical observations point towards a decelerated expansion of the universe just prior to the present acceleration, during a radiation dominated era and matter dominated era just before, in order to allow nucleosynthesis and the formation of galaxies. This is realized by a signature flip of the deceleration parameter from negative to positive, $q = -\left(\ddot{a}/a \right)/\left(\dot{a}^2/a^2 \right) > 0$ \citep{riess, paddy}, beyond a certain value of the redshift $z$. The most plausible way to describe this is to consider that the so-called Dark Energy component, the driver of late-time acceleration, somehow remains subdued during an earlier epoch and dominates in the late-time. The most simple model widely considered in this regard are the cosmological constant $\Lambda$. On the other hand, time evolving scalar field models known as the quintessence are popular as well, where the self-interaction potential of the scalar field dominates over the kinetic term at late times to generate the effective negative pressure. However, it is indeed more appropriate to employ a scalar field which has it's genesis rooted in the realm of field theory. This motivation leads one to revisit the so-called Scalar-Tensor generalizations of GR, of which Brans-Dicke theory is the prototype. Keeping in mind the constraints over $\omega$, one often considers generalized setups of BD theory, for example by taking $\omega$ as a function of the BD field \citep{berg, wagoner, nordt, bark, vdb}. Different aspects of these generalizations including their application in cosmology are well-studied in literature \citep{koli, holden, santos, bert, nandan}. Non-trivial generalizations such as interaction between a dark matter distribution and the BD scalar field has also received some attention \citep{sudiptadigrg, clifton1, sudiptadi}. These considerations carry a remarkable similarity in setup with the idea of Chameleon scalar field cosmology where the scalar field couples to ordinary matter and has a local matter density dependent mass \citep{khoury, mota, dascora, motashaw1, motashaw2}. These models can also serve as a probable way to look into dark energy-dark matter interaction and to seek for a solution of the cosmic {\it `coincidence problem'}, as an extension of existing novel multiple scalar models \citep{saa, saa1, elizalde}.  \\

We explore the BD field profile and the strength of it's interactions in an extended BD theory where the standard action is extended by including a new massive scalar field. The additional field carries resemblance with a Higgs boson, as in the definition of it's interaction potential. The BD field $\psi$ interacts with curvature in the standard manner and the Higgs field $\phi$ interacts with gravity derivatively as well as non-derivatively \citep{amendolan, capopo}. No self-interaction potential for the BD field is assigned. This is indeed motivated by the subtle overlap of interests in between Particle Physics and GR, which has also led physicists towards the Higgs inspired models of early inflation \citep{alexander1, alexander2, bez, germani, masina, tsuji} as well as Higgs cosmology consistent with astrophysical observations and physics in LHC \citep{atkins, xian, wegner1, wegner2}. The extended theory considered in this manuscript is primarily inspired from a recent self-consistent derivation of Higgs potential. The proof is valid for simple power-law cosmological solutions \citep{sola} which are asymptotic states of a more general phase of cosmic evolution \citep{amendoo}. The underlying philosophy also inspires an idea of cosmic acceleration driven by dynamical dark energy in a Higgs generalized BD setup \citep{sola1}, or the running vacuum $\Lambda$CDM model \citep{sola2}. The basis of these ideas has already been discussed at quite a length, both at the theoretical and phenomenological level \citep{sola3, sola4}. We mean to expand the understanding a bit through a simple motivation. It is not possible to describe the deceleration-to-acceleration transition of the universe at late times using a power law cosmological solution, since it produces a constant deceleration parameter. We consider a more general cosmological solution that can fit well with astrophysical observations. We find such a solution through a cosmological reconstruction and show that the universe has a smooth transition from a decelerated era into an epoch of accelerated expansion around the observationally consistent redshift of transition, $z_{t} < 1$ \citep{farooq, purbasoumya}. \\

A reconstruction or a method of reverse engineering of the cosmic evolution has become an increasingly popular methodology of late. The popularity is motivated mainly from the fact that one can avoid writing an exact solution of the highly non-linear field equations. Primarily, a reconstruction scheme finds it's origin from comparison with observational data or some phenomenological ansatz \citep{peebles, copeland}. Broadly the popular approaches of cosmological reconstruction can be divided in two groups; Parametric approaches \citep{chen, ryan}, based on some pre-assumed parametric form of one or more cosmological quantity and Non-parametric approaches, involving statistical methods of comparing large sets of observational data, for instance, principle component analysis \citep{critt, clarkson, ishida, amendola}, gaussian process \citep{hols, seikel, shafi}. Reconstruction from kinematic approach is also quite well posed and both parametric and non-parametric reconstruction using purely kinematical quantities are receiving increasing attention over the past few decades \citep{berns, visser, catt, aviles, busti, duna, mukherjee1, mukherjee2, mukherjee3}. We work out an analytical reconstruction using the cosmological statefinder parameter {\bf \it 's'}. The statefinder parameter is written as a dimensionless combination of the cosmic jerk parameter and the deceleration parameter \citep{vsahni, alam, evans}. Assuming that the statefinder parameter remains a constant during the late-time evolution of the universe, we solve for the Hubble as a function of redshift $z$. We compare the Hubble free luminosity distance measurement data with the theoretically calculated spectrum and estimate the model parameters. We study the evolution of the deceleration parameter for the best fit model and check whether the required signature flip in the evolution can be realized. We reconstruct and study the interaction functions of the scalar fields, namely the BD scalar $\psi$ and the Higgs scalar $\phi$, as a function of redshift. This allows us to understand the structure of the extended BD theory that can support the required signature flip of the deceleration parameter, or in other words, can describe a viable late-time cosmology. As it turns out, both of the scalar fields and their interaction play a crucial part in switching on or off the cosmic acceleration or deceleration in different epoch of expansion of the universe. The gravitational interaction in the theory is stronger in earlier epoch where Hubble rate is low. With an accelerated expansion Hubble increases and the interaction weakens, making the extended Brans-Dicke-Higgs theory considered here an asymptotically free theory.   \\

In Section $2$ we introduce the basc action, mathematical motivation and the field equations of the theory. Section $3$ introduces the detailed work out of the cosmological reconstruction from a constant statefinder parameter. We also include brief discssions on the study of relevant cosmological quantities like the effective equation of state parameter, matter density contrast and thermodynamic equilibrium of the universe in Section $3$. Section $4$ contains numerical reconstruction of the extended BD theory lagrangian, in the form of the evolution of the scalars and their interaction as a function of redshift. We conclude the manuscript in Sect. $5$.

\section{A Brans-Dicke-Higgs Scalar Extended Theory}
The theory we consider is a generalized two-scalar BD setup. The standard BD scalar field is written as $\psi$. A second scalar field $\phi$ is included in the action that has derivative as well as non-derivative interaction with gravity. The action is written as
\begin{eqnarray}\nonumber
&& S = \int d^{4}x\sqrt{-g}\Bigg[\frac{1}{2}R\psi - \frac{\omega}{2\psi}g^{\mu\nu}\partial_{\nu}\psi\partial_{\mu}\psi + U(\phi , \psi) \\&& 
+ \xi R\phi^{2} - \frac{1}{2} g^{\mu\nu}\partial_{\mu}\phi\,\partial_{\nu}\phi + \frac{1}{\phi^{2}}S_{\mu\nu}\partial^{\mu}\phi\partial^{\nu}\phi - V(\phi)\Bigg].  \label{eq:SMBDaction}
\end{eqnarray}
There is no self-interaction potential of the BD field $\psi$. Since we are interested in the cosmology described by this setup when the scalar fields dominate over all the other field sources, we do not add any matter energy momentum tensor. In order to describe the time evolution of the current universe, it is necessary to assign some form of pressureless matter as well as a dark energy fluid in the lagrangian of the theory. In the present setup we keep the onus on two scalar fields and their interaction profiles to fill in for an effective fluid description driving the acceleration. This is additionally motivated from a vintage proof that a spherically symmetric scalar field with self-interaction can mimic a dust matter evolution \citep{goncamoss}. Therefore, there remains a prospect in this two-scalar setup that one of the scalar fields can play the role of a scalar field dark matter. Moreover, one can argue and draw some similarities between the BD scalar itself playing the role of dark energy without the need of a cosmological constant or a quintessence matter \citep{pavon}. We motivate ourselves further and consider an interacting dark matter-dark energy scenario using an interacting two-field lagrangian since there are no theoretical arguments forbidding such an interaction, nor there is any sufficient observational results to rule it out. Moreover, the fact that the present energy densities of dark energy and matter are observed to be the same order of magnitude, suggests a connection between them. An intimate connection between these components is also naturally inspired by generalized models of unification \citep{dmdeint1, dmdeint2, dmdeint3, interdmde1, interdmde2, interdmde3}. The interaction potential can be written in the form of a dark matter mass and this can lead one to an `interacting quintessence' scenario \citep{bertolamiinter}. In a nutshell, this constitutes a classic way to write the acceleration/deceleration of our universe as a combined effort of the dark components of the universe with hints from fundamental physics models. The extended theory and the interaction profiles considered here are a straightforward generalization of the work of \cite{sola}, however, we do not specify any form of the interaction $U(\phi,\psi)$ between the two scalars at the outset. We take both $\psi$ and $\phi$ as spatially homogeneous fields and also keep in mind that the BD scalar field $\psi$ in standard or generalized BD-theory is expected to be a slowly varying function of time \citep{Damour} in order to satisfy the Weak Equivalence Principle. By proceeding in this way our framework may have additional parameters over the parameters of the action. For instance the ones in the Higgs potential, in the end remain unrelated to those in the action and are treated as free parameters albeit with phenomenological constraints. The tensorial term $S_{\mu\nu}$ allows generalized derivative interaction of $\phi$ with gravity. 
\begin{equation}\label{eq:SmunuTensor}
S_{\mu\nu}\equiv\varsigma R_{\mu\nu}-\frac{\theta}{2}g_{\mu\nu}R. 
\end{equation}
The non-minimal derivative interaction in the action is inspired from models that produce inflationary attractor solutions \citep{amendolan, capopo} in cosmology. These interactions are also found in other novel areas of physics, for example, in the action of scalar quantum electrodynamics a derivative coupling between the electro-magnetic vector and the scalar field is required to satisfy the $U(1)$ gauge-invariance of the theory. A generalized derivative couplings also leads to the scalar formulation of the hypothesis that $G$ is a function of the energy density of the gravitational field source \citep{amendolan}. We choose $\theta \neq \varsigma$, since the equality simply makes $S_{\mu\nu}$ proportional to the Einstein tensor $G_{\mu\nu}$. We use a rescaled BD-field $\psi \sim 8\pi\psi$ for our calculations in natural units. With this, the effective gravitational coupling is inversely proportional to the BD scalar field
\begin{equation}\label{eq:Geff}
G_{eff}(t) = \frac{1}{8\pi\psi(t)}.
\end{equation}
The BD scalar field is of the order of Planck Mass squared. Around present time $M_P = 1/\sqrt{8\pi G} \simeq 2.43\times 10^{18}$ GeV. In natural units, the scalar $\psi$ has a dimension of mass squared and the additional scalar $\phi$ has mass dimension $1$. The BD-parameter $\omega$ is kept a constant in order to keep the modified theory as close to the standard setup as possible. $\xi$ is a dimensionless parameter characterizing the interaction of $\phi$ with gravity. Ordinarily, in the limit $\omega\to\infty$ the action in Eq. (\ref{eq:SMBDaction}) can reduce to GR if the generalized couplings are zero. However, we keep all the couplings non-zero and discuss the compatibility with observational data. We choose that the self-interaction of $\phi$, $V(\phi)$ behaves as a Higgs Potential at the outset, $V(\phi) \sim \alpha \phi^2 + \beta \phi^4$. The philosophical relevance and versatility of the choice of the potential shall be discussed in more details in Sec. $4$. The field equations of the theory are obtained by varying the action with respect to metric and the scalar fields.

\begin{eqnarray}\nonumber\label{VariationMetric}
&& (G_{\mu\nu} + D_{\mu\nu})\psi + \frac{\omega}{2\psi}g_{\mu\nu}(\nabla\psi)^2 - \frac{\omega}{\psi}\nabla_{\nu}\psi\nabla_{\mu}\psi - g_{\mu\nu} \\&&\nonumber
 U(\phi, \psi) - \nabla_{\nu}\phi\nabla_{\mu}\phi + \frac{1}{2}g_{\mu\nu}\nabla_{\alpha}\phi\nabla^{\alpha}\phi + 2\xi(G_{\mu\nu}  + D_{\mu\nu}) \\&&\nonumber 
\phi^{2} + 2\varsigma \Bigg\lbrace \frac{1}{2}g_{\mu\nu}\nabla_{\beta}\nabla_{\alpha}\Big(\phi^{-2}\nabla^{\alpha}\phi\nabla^{\beta}\phi\Big) + R_{\mu\alpha}(\phi^{-2}\\&&\nonumber
\nabla^{\alpha}\phi\nabla_{\nu}\phi) + R_{\nu\alpha}\Big(\phi^{-2}\nabla^{\alpha}\phi\nabla_{\mu}\phi\Big) - \frac{1}{2}g_{\mu\nu}R_{\alpha\beta}\Big(\phi^{-2}\\&& \nonumber
\nabla^{\alpha}\phi\nabla^{\beta}\phi\Big) - \frac{1}{2}\Big(\nabla_{\mu}\nabla_{\beta} (\phi^{-2}\nabla_{\nu}\phi\nabla^{\beta}\phi) + \nabla_{\nu}\nabla_{\beta}(\phi^{-2}\\&&\nonumber
\nabla_{\mu}\phi\nabla^{\beta}\phi)\Big) + \frac{1}{2}\square(\phi^{-2}\nabla_{\mu}\phi\nabla_{\nu}\phi\Bigg) \Bigg\rbrace -\theta\Bigg\lbrace (G_{\mu\nu} \\&&\nonumber
+ D_{\mu\nu})(\phi^{-2}\nabla_{\alpha}\phi\nabla^{\alpha}\phi) + R (\phi^{-2}\nabla_{\mu}\phi\nabla_{\nu}\phi)\Bigg\rbrace \\&&
+ g_{\mu\nu}V(\phi) = 0, \\&& 
D_{\mu\nu}\equiv g_{\mu\nu}\square-\nabla_{\mu}\nabla_{\nu}.
\end{eqnarray}
The evolution equation of the scalar fields are written as
\begin{eqnarray}\nonumber
&&\Box\phi + \frac{2}{\phi^{3}}S_{\mu\nu}\nabla^{\mu}\phi\nabla^{\nu}\phi - \frac{2}{\phi^{2}}\Big\lbrace(\nabla^{\mu} S_{\mu\nu})\nabla^{\nu}\phi \\&&
+ S_{\mu\nu}(\nabla^{\mu} \nabla^{\nu}\phi)\Big\rbrace + \frac{\partial U}{\partial \phi} + 2\xi R \phi-\frac{dV}{d\phi} = 0, \label{Variationphi}
\end{eqnarray}
and
\begin{equation}
\Box\psi-\frac{1}{2\psi}(\nabla\psi)^2+\frac{\psi}{2\omega}\,R + \frac{\psi}{\omega}\frac{\partial U}{\partial \psi}=0.
\label{Variationpsi}
\end{equation}
In order to generate and study cosmological solution, we take a spatially flat Friedmann-Robertson-Walker (FRW) metric written as
\begin{equation}
ds^{2} = -dt^{2} + a^{2}(t)\left(dr^{2} + r^{2}d\Omega^{2}\right). \label{eq:frwmetric}
\end{equation}

For an FRW metric, the field equations become
\begin{eqnarray}\nonumber
&& {{3}H^{2}\psi}+{{3}H{\dot{\psi}}}-\frac{\omega}{2}\frac{\dot{\psi}^{2}}{\psi} + U(\phi, \psi) -\frac{1}{2}\dot{\phi}^{2}-V(\phi) \\&&\nonumber
+ 6\xi H^{2}\phi^{2} + 12\xi H \dot{\phi}\phi - {9}\theta H^{2}\frac{\dot{\phi}^{2}}{\phi^{2}} -  6(\theta-\varsigma)\dot{H}\frac{\dot{\phi}^{2}}{\phi^{2}} \\&&
+ 6(\theta-\varsigma)H\frac{\dot{\phi}\ddot{\phi}}{\phi^{2}} - 6(\theta -\varsigma){H}\frac{\dot{\phi}^{3}}{\phi^{3}} = 0,
 \label{eq:EoM-metric}
\end{eqnarray}

\begin{eqnarray}\nonumber
&&\ddot{\phi}+3 H \dot{\phi}-12\xi\dot{H}\phi - 24\xi H^{2}\phi + \frac{d V}{d\phi} + 6 (2\theta - \varsigma) H^{2} \\&& \nonumber
\Bigg(\frac{\ddot{\phi}}{\phi^{2}} - \frac{\dot{\phi}^{2}}{\phi^{3}}\Bigg) + 18 (2\theta -\varsigma){H}^{3}\frac{\dot{\phi}}{\phi^{2}} + 6 (7\theta - 5\varsigma) H \dot{H}\frac{\dot{\phi}}{\phi^{2}} \\&&
+ 6 (\theta - \varsigma)\ddot{H}\frac{\dot{\phi}}{\phi^{2}} + 6(\theta -\varsigma)\dot{H}\Big(\frac{\ddot{\phi}}{\phi^{2}}-\frac{\dot{\phi}^{2}}{\phi^{3}}\Big) - \frac{\partial U}{\partial \phi} = 0, \label{eq:EoM-phi}
\end{eqnarray}
and
\begin{equation}
3\dot{H}+6{H}^{2} - \omega \frac{\ddot{\psi}}{\psi}+\frac{ \omega}{2} \frac{\dot{\psi}^{2}}{{\psi}^{2}}-3H\omega\frac{\dot{\psi}}{\psi}+ \frac{\partial U}{\partial \psi} = 0. \label{eq:EoM-psi}
\end{equation}
Using the field equations, one can prove that the effective Einstein tensor equivalent $G^{eff}_{\mu\nu}$ is covariantly conserved, indicating that the system does not avail the possibility of spontaneous matter creation or annihilation. The field Eqs. (\ref{VariationMetric}), (\ref{Variationphi}) and (\ref{Variationpsi}) are similar to the equations derived by \cite{sola}, except the fact that the cross interaction term $U$ has no preassigned functional form here (It can be checked that for $U\sim \psi\phi^2$ one recovers the results of \cite{sola}). Before going further we emphasize again, that our goal is to look into the evolution of the interaction functions that can support a preassumed Higgs scalar and a smooth deceleration to acceleration transition of the time evolving universe. A Higgs potential comes as an artifact of the extended theory itself \citep{sola} which can be proved using a power-law cosmological solution. However, a smooth signature flip in deceleration parameter is not possible with a power law cosmological solution (even with an $U\sim \psi\phi^2$), which yields a constant deceleration parameter. Keeping this in mind we tweak the lagrangian enough to allow an extension in the form of a generalized $U$. We find out the profile of $U$ that supports a pre-assumed Higgs potential in the lagrangian and a smooth deceleration to acceleration transition of the universe.

\section{Reconstruction from The Statefinder Parameter}\label{s1}
Finding an exact solution of the system of equations is no doubt extremely non-trivial and therefore we resort to a cosmological reconstruction. The cosmic statefinder parameter is a kinematic quantity, written as a dimensionless combination of the Hubble parameter $H(z)$ and it's derivatives. The Hubble and the Deceleration parameter involve the first and the second derivative of the scale factor, 
\begin{eqnarray}
H & = & \frac{\dot{a}}{a} \label{eq:1.1} \\
q & = & -\frac{\ddot{a}a}{\dot{a}^2} = -\frac{\dot H}{H^2}-1, \label{eq:eq1.2}
\end{eqnarray}
where $a$ is the scale factor and by a dot we denote derivative with respect to cosmic time. The statefinder parameters are defined as \citep{vsahni, alam}.
\begin{eqnarray}
r & = & \frac{\stackrel{\bf{...}}{a}}{aH^3} 
=\frac{\ddot H}{H^3}+3\frac{\dot H}{H^2}+1\label{eq:eq1.3} \\
s & = & \frac{r-1}{3\left(q-\frac{1}{2}\right)}. \label{eq:eq1.4}
\end{eqnarray}

Writing $1+z = \frac{1}{a} \equiv x$, $H' = \frac{dH}{dx}$ and $a'= -a^2$ we can express the above quantities as a function of $x$ 
\begin{eqnarray}
&&\dot{H} = -H'H/a, \\&&
q(x) = \frac{H'}{H}x-1. \label{eq:eq1.9} \\&&
r(x) = 1-2\frac{H'}{H}x + \left(\frac{H'^2}{H^2}+\frac{H''}{H}\right)x^2. \label{eq:eq1.14}
\end{eqnarray}

The statefinder $s(x)$ is written putting Eqs. (\ref{eq:eq1.9}) and (\ref{eq:eq1.14}) in Eq. (\ref{eq:eq1.4}),

\begin{equation}
s(x) = \frac{-2x\frac{H'}{H} + \left(\frac{H'^2}{H^2}+\frac{H''}{H}\right)x^2}{3\left(\frac{H'}{H}x - \frac{3}{2} \right)}.
\end{equation}

We re-write this as a second order differential equation for the Hubble,
\begin{equation}\label{mastereq1}
3 s \frac{H'}{H} x - \frac{9s}{2} = -2\frac{H'}{H}x + \frac{H'^2}{H^2}x^2 + \frac{H''}{H}x^2.
\end{equation}

Our assumption is that the statefinder parameter $s(x)$ can be rendered as a constant during the deceleration to acceleration transition of the universe at late-times. This can be thought of as a special case of more generalized scenario where $s(x)$ is parameterized as a function of redshift. We assign the value $s = \delta - \frac{2}{3}$. This infact serves as an alternative way to look into extended scalar-tensor theories through analytical solutions. While there are a plethora of research works on cosmological solutions and analysis of observational data in similar extended theories \citep{sola2, sola3, sola4}, comparisons with these novel studies are not really our sole purpose. Now, we solve Eq. (\ref{mastereq1}) and write Hubble as a function of $x$. The equation can be recast as

\begin{equation}\label{mastereq2}
\frac{H''}{H} + \frac{H'^2}{H^2} - \frac{3\delta}{x}\frac{H'}{H} - \frac{9}{2}\left(\delta - \frac{2}{3}\right)\frac{1}{x^2} = 0,
\end{equation}

with the solution 
\begin{eqnarray}\nonumber\label{hubblex}
&& H(x) = C_{2}x^{\frac{1}{4}\left[1+3\delta- \left\lbrace 1 + 6\delta + 9\delta^2 + 36\left(\delta - \frac{2}{3}\right)\right\rbrace^{1/2}\right]} \\&&
\left[C_{1} + x^{\left\lbrace 1 + 6\delta + 9\delta^2 + 36\left(\delta - \frac{2}{3}\right)\right\rbrace^{1/2}}\right]^{1/2}.
\end{eqnarray}

$C_{1}$ and $C_{2}$ are constants of integration. Straightaway Eq. (\ref{hubblex}) produces a constraint over the parameter $\delta$. For a real evolution one must ensure that all the powers of $x$ on the RHS are real. This means $\left\lbrace 1 + 6\delta + 9\delta^2 + 36\left(\delta - \frac{2}{3}\right)\right\rbrace > 0$. This is simplified into
\begin{equation}
\delta^2 + \frac{14}{3}\delta - \frac{23}{9} > 0.
\end{equation} 

The Model in Eq. (\ref{hubblex}) gives the deceleration parameter as a function of redshift as is clear from the definition in Eq. (\ref{eq:eq1.9}). This can show a possibility for signature flip in the expression of $q$. In a bid to look into this from a more realistic angle, we try and estimate the model parameters from observational data. We use three different data sets and draw the confidence contours in comparison with the best fit values in the parameter space. The primary data set comes from the Supernova distance modulus-luminosity distance measurement data. The other two are from the Hubble parameter measurements (OHD) and the Baryon Acoustic Oscillation (BAO) data. To compare the reconstructed set of cosmological parameters with observational data, we express all the relevant expressions in their dimensionless form, as a function of redshift $z$ which is related to the scale factor through
\begin{equation}
(1 + z) = \frac{a_0}{a(t)}.
\end{equation}
$a_0$ is the scale factor at current epoch. Using this we can change the arguement into redshift from cosmic time. Using Eq. (\ref{hubblex}) and $1+z \equiv x$ we rewrite the parametric expression for the Hubble as
\begin{eqnarray}\nonumber\label{hubblez}
&& H(z) = \\&&\nonumber
\frac{H_{0}}{(1+C_{1})^{1/2}}(1+z)^{\frac{1}{4}\left[1+3\delta- \left\lbrace 1 + 6\delta + 9\delta^2 + 36\left(\delta - \frac{2}{3}\right)\right\rbrace^{1/2}\right]} \\&&
\left[C_{1} + (1+z)^{\left\lbrace 1 + 6\delta + 9\delta^2 + 36\left(\delta - \frac{2}{3}\right)\right\rbrace^{1/2}}\right]^{1/2}.
\end{eqnarray}
$H_0$ is the current value of Hubble parameter whose dimension is km $\mbox{Mpc}^{-1}$ $\mbox{sec}^{-1}$. The Hubble function can be written in a dimensionless form by scaling $H_0$ by $100$ km $\mbox{Mpc}^{-1}$ $\mbox{sec}^{-1}$. The reduced Hubble function is written as
\begin{equation}
h(z) = \frac{H(z)}{H_0} = \frac{H(z)}{100\times h_0}.
\end{equation}
Essentially, we constraint the dimensionless parameter $h_{0} = H_{0}/100 km\mbox{Mpc}^{-1} \mbox{sec}^{-1}$, the deceleration parameter at the current epoch and the redshift of transition when the universe undergoes a transition from deceleration into acceleration. \\

\begin{figure}
\begin{center}
\includegraphics[angle=0, width=0.52\textwidth]{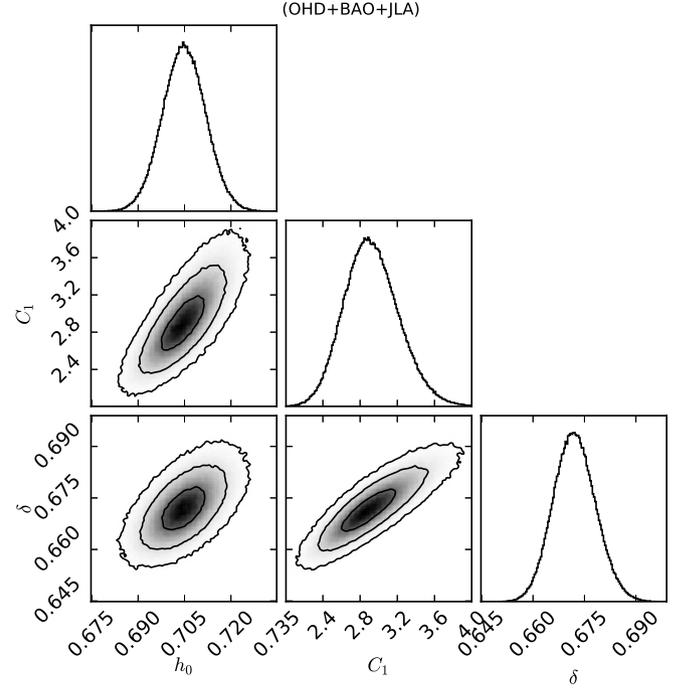}
\caption{Confidence contours on the parameter space and the marginalized likelihood function, obtained from the combined analysis of OHD+JLA+BAO, alongwith the associated 1$\sigma$, 2$\sigma$ confidence contours.}
\label{Modelcontour}
\end{center}
\end{figure}

The datasets taken into account for the study of cosmological constraints are
\begin{itemize}
\item {Supernova distance modulus data from the Joint Light-Curve Analysis \citep{jla} of $SDSS-II$ and $SNLS$ collaborations.}
\item {Hubble parameter estimation data \citep{ohd1, ohd2, ohd3, ohd4, ohd5, ohd6, delu} and the measurement of $H_0$ by \cite{planck}.}
\item {The Baryon Acoustic Oscillations (BAO) data, given by measurements of $\frac{r_s(z_d)}{D_v(z_{BAO})}$, where $r_s(z_d)$ is the sound horizon, $z_d$ is called the photon drag epoch and $D_v$ is dilation at the measurement redshift. We use data from three measurements at different redshift ($6dF$ $Galaxy$ $Survey$ \citep{6dF}, BOSS LOWZ and BOSS CMASS \citep{bossanderson}). In standard manner, the BAO data is scaled using data from Planck \cite{planck}.}
\end{itemize}

We estimate the uncertainty and the likelihood of possible parameter values, using a Markov Chain Monte Carlo numerical simulation (MCMC). The code used in this manuscript is an execution of the numerical code in python, called the {\it `emcee'} \citep{emcee}. In Fig. \ref{Modelcontour}, we show the estimated parameters of the model, using confidence contours and likelihood functional analysis. The plots show that the model parameters have a positive correlation between themselves. The best-fit values of the model parameters alongwith 1$\sigma$ estimated errors are given in Table \ref{resulttable}. The best fit parameters suggest that the present value of Hubble parameter found from the reconstructed model is well consistent with observational data. The best fit value of the statefinder parameter $s = \delta - \frac{2}{3}$ is estimated to be in the range $0.011 \geq s \geq -0.001$.\\

\begin{table}
\caption{{\small The parameter values and the associated 1$\sigma$ uncertainty of the parameters, obtained from the analysis with different combinations of the data sets.}}\label{resulttable}
\begin{tabular*}{\columnwidth}{@{\extracolsep{\fill}}lrrrrl@{}}
\hline
 & \multicolumn{1}{c}{$h_0$} & \multicolumn{1}{c}{$C_{1}$} & \multicolumn{1}{c}{$\delta$} \\
\hline
$OHD+JLA+BAO$ 	  & $0.705^{+0.007}_{-0.007}$ &$2.913^{+0.303}_{-0.274}$ & $0.672^{+0.006}_{-0.006}$ &\\
\hline
\end{tabular*}
\end{table}

\begin{figure}
\begin{center}
\includegraphics[angle=0, width=0.40\textwidth]{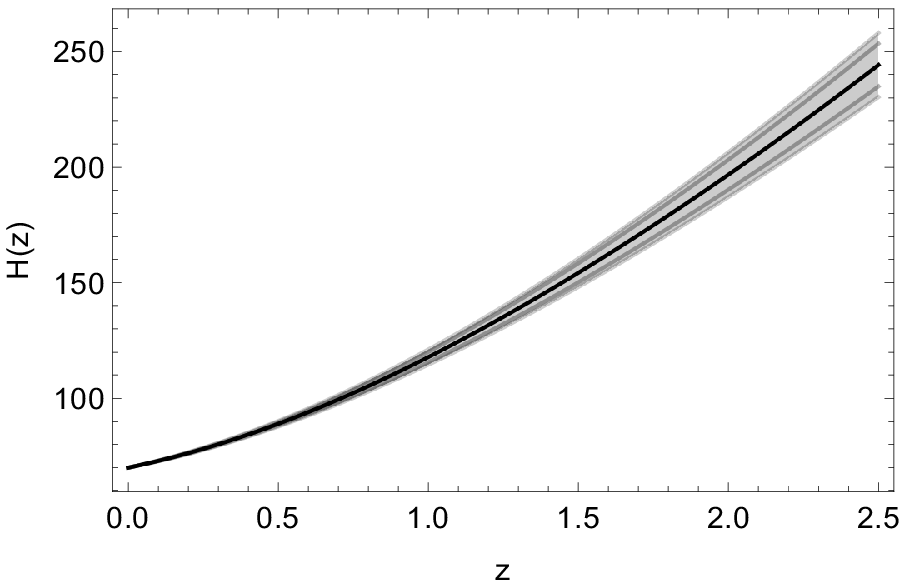}
\includegraphics[angle=0, width=0.40\textwidth]{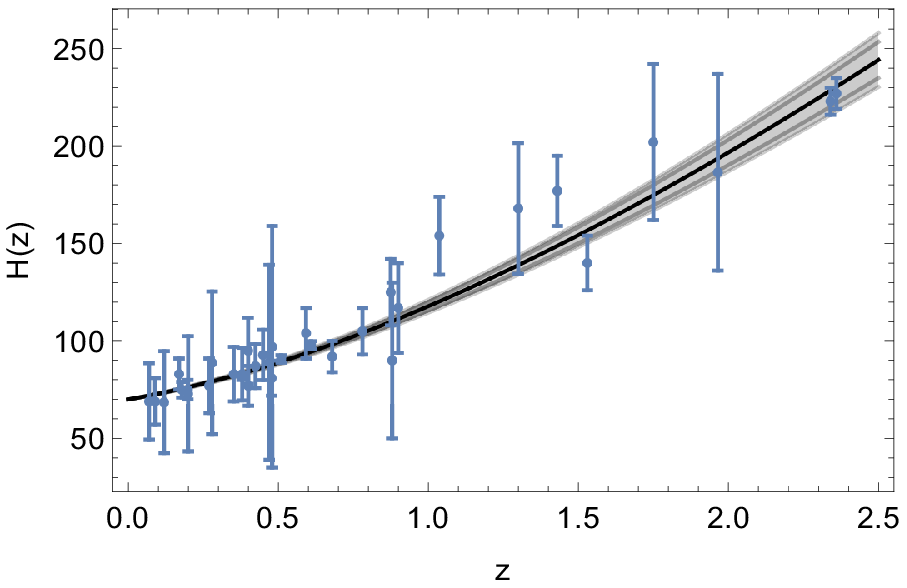}
\caption{Plot of the reconstructed Hubble parameter $H(z)$ as a function of redshift alongwith observational data points. The thick black line is for best fit parameter values and the gray regions are for associated 2$\sigma$ and 3$\sigma$ confidence regions.}
\label{Hz_data}
\end{center}
\end{figure}

Relevant cosmological quantities can be plotted as a function of redshift for the set of parameters giving best fit as well as for uncertainty regions. Fig. \ref{Hz_data} show that evolution of $H(z)$ for the reconstructed model are well consistent with the observational data in the low redshift regime. A plot in the 1$\sigma$ confidence region almost coincides with the best fit parameter plot and therefore we show only 2$\sigma$ and 3$\sigma$ confidence region plots. Moreover, the best fit value of the Hubble parameter around $z \sim 0$, predicted by the present model is very close to astrophysical observations \citep{riess2018}.

\begin{figure}
\begin{center}
\includegraphics[width=0.40\textwidth]{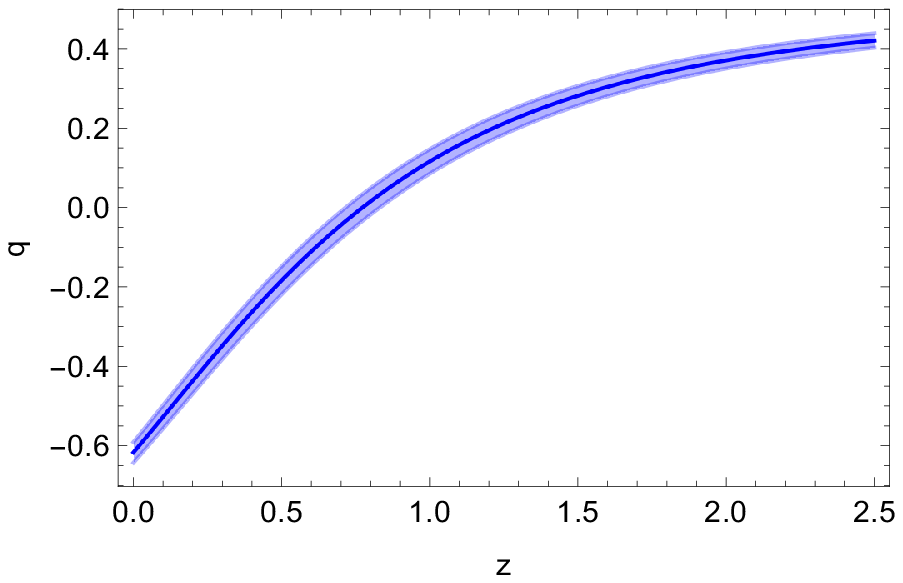}
\includegraphics[width=0.40\textwidth]{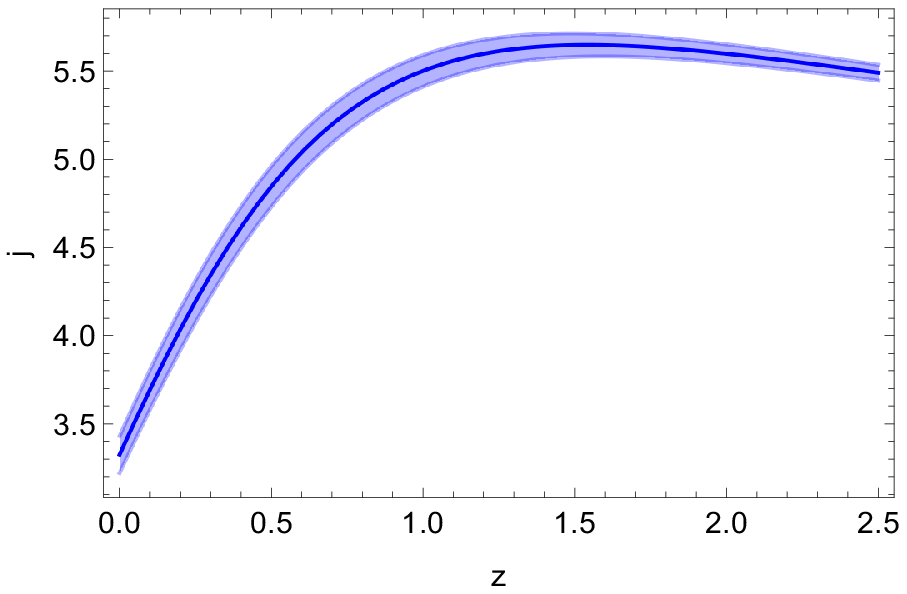}
\caption{Evolution of the deceleration parameter $q(z)$ (top graph) and the jerk parameter $j(z)$ (bottom graph) for the best fit parameter values (bold blue curve) and the associated 3$\sigma$ confidence region (faded blue curve).}
\label{kinematic_parameters}
\end{center}
\end{figure}

We plot the evolution of the deceleration parameter $q(z)$ and the jerk parameter $j(z)$ as a function of redshift in Fig. \ref{kinematic_parameters}. The plots are for the best fit parameter values (bold blue curve) and the associated 3$\sigma$ confidence region (faded blue curve). The value of the deceleration parameter at the present epoch ($z \sim 0$) is close to $-0.62$ which agrees quite well with observational data and keepin in mind a corresponding $\Lambda$CDM cosmology. There is a transition in the signature of $q(z)$ from a decelerated phase into an accelerated phase of expansion. The transition redshift $z_{t} < 1$ is also consistent with direct observational results \citep{riessobs, farooq}. As the deceleration parameter shows a non-trivial evolution with respect to redshift, it is very important to investigate the next order derivative of the scale factor. The lower curve of Fig. \ref{kinematic_parameters} shows a non-trivial evolution of jerk parameter as a function of redshift and the present value shows a clear departure for $\Lambda$CDM. \\

Since the entire scheme of this reconstruction depends only on one kinematical quantity, the resulting structure is independent of any assumption over the dynamics of dark energy or any specific theory of gravity. Therefore, if the theory under consideration is standard GR, we can also plot the evolution of the equation of state parameter $w_{eff}$ of a preassumed energy momentum distribution as

\begin{figure}
\begin{center}
\includegraphics[width=0.40\textwidth]{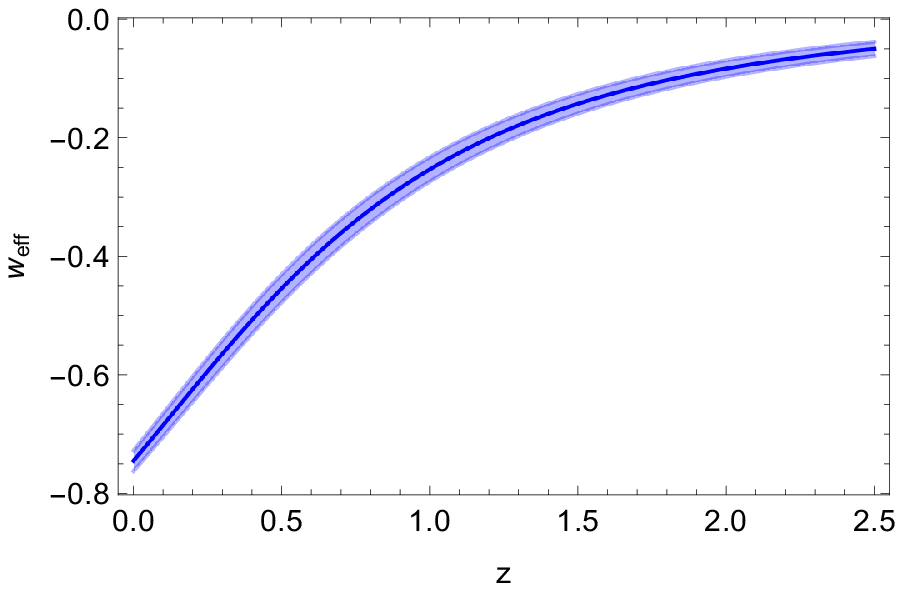}
\includegraphics[width=0.40\textwidth]{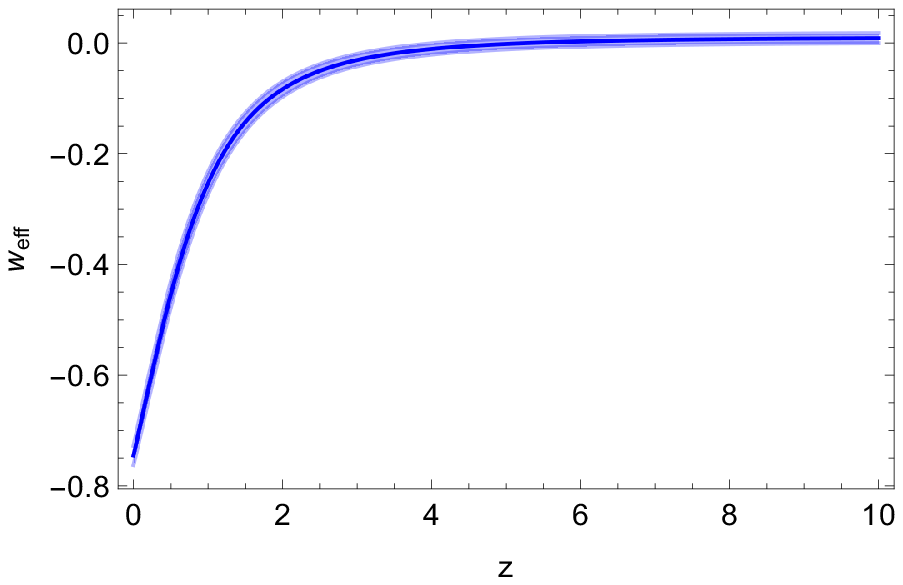}
\caption{Evolution of equation of state as a function of redshift for the best fit parameter values (bold blue curve) and the associated 3$\sigma$ confidence region (faded blue curve).}
\label{eosfig}
\end{center}
\end{figure}

\begin{equation}
w_{eff}=\frac{p_{tot}}{\rho_{tot}}.
\end{equation}
We can also relate it to the Hubble function $H(z)$ through
\begin{eqnarray}\nonumber
&&\frac{\rho_{tot}}{\rho_{c0}} = \frac{H^2(z)}{H^2_0}, \\&&\nonumber
\frac{p_{tot}}{\rho_{c0}}=-\frac{H^2(z)}{H^2_0}+\frac{2}{3}\frac{(1+z)H(z)H'(z)}{H^2_0}, \\&&\nonumber
\rho_{c0}=3H^2_0/8\pi G.
\end{eqnarray}

\begin{figure}
\begin{center}
\includegraphics[width=0.40\textwidth]{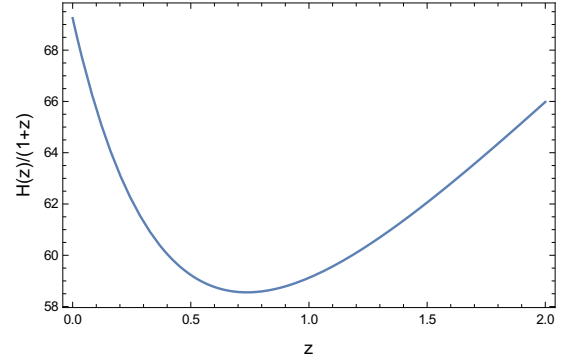}
\caption{Evolution of $\frac{H(z)}{(1+z)}$ for the best fit parameter values $h_{0} \sim 0.71$, $C_{1} \sim 2.913$ and $\delta \sim 0.67$.}
\label{other_parameters}
\end{center}
\end{figure}

$\rho_{c0}$ is the critical density. The graph on the top panel of Fig. \ref{eosfig} shows $w_{eff}$ as a function of redshift. The plot clearly shows that for a low $z$, $w_{eff} < 0$. $w_{eff}$ increases with $z$ towards a zero value. This indicates an effective negative pressure experienced by the universe during the present epoch leading to the accelerated expansion. In recent past the universe was in a dust dominated epoch where $w_{eff} \sim 0$ as shown in the bottom panel of Fig. \ref{eosfig}. Therefore, irrespective the evolution of jerk parameter showing a clear departure from $\Lambda$CDM, these plots seem to echo some of the behavior. This is further confirmed by the evolution of $H(z)/(1+z)$ as shown by the graph in Fig. \ref{other_parameters}. Apart from a very slight deviation at high redshift, the curve closely follows a behavior suggestive of $\Lambda$CDM cosmology.

With cosmic acceleration, a growth of matter over-density is also expected. This can generate a `matter density contrast', $\delta_m = \frac{\delta\rho_m}{\rho_m}$. For a perfect fluid in GR setup, let us say that the background matter density is $\rho_m$ which is homogeneous. Deviation from the background density is written as $\delta\rho_m$. While $\delta_m$ can show a non-linear evolution around locally over-dense regions of cosmos, during a cosmic expansion it is safe to assume that $\delta_m$ behaves in an approximately linear manner. The time evolution of $\delta_m$ is a key factor behind the eventual structure formation of the universe and at linear level is written as
\begin{equation}\label{delm_eq}
\ddot{\delta}_m + 2H\dot{\delta}_m = 4\pi G\rho_m\delta_m.
\end{equation}

\begin{figure}
\begin{center}
\includegraphics[width=0.40\textwidth]{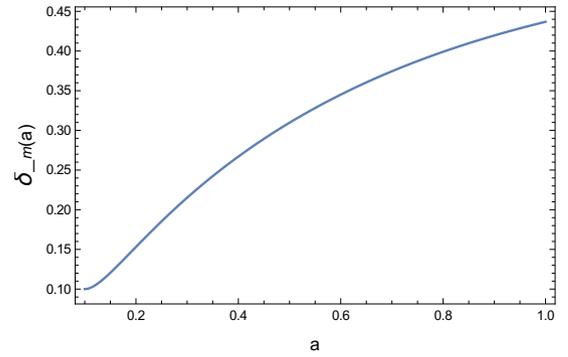}
\caption{Evolution of growth of matter over-density $\delta_m$ as a function of scale factor for the best fit parameter values $h_{0} \sim 0.71$, $C_{1} \sim 2.913$ and $\delta \sim 0.67$.}
\label{overdensity}
\end{center}
\end{figure}

One can study $\delta_m$ numerically, as a function of scale factor $a$ from Eq. (\ref{delm_eq}). For a perfect fluid, one can write $\rho_{m} = \frac{\rho_{0}}{a^3}$ and solve the equation with initial conditions $a_i = 0.01$ and the initial values are fixed as, $\delta_m(a_i) = 0.01$ and $\dot{\delta}_m(a_i) = 0$. The evolution of $\delta_m$ vs $a$ for the best fit parameter values $h_{0} \sim 0.71$, $C_{1} \sim 2.913$ and $\delta \sim 0.67$ is shown in Fig. \ref{overdensity} and the evolution is quite similar to the corresponding $\Lambda$CDM cosmology. \\

We assume that the cosmological system under consideration obeys the laws of blackhole thermodynamics and is surrounded by an apparent horizon or simulteneously, the Hubble horizon for a spatially flat FLRW geometry \citep{gibbon, jacobson, bak}. The total entropy of the universe can be seen as an aggregate of fluid entropy components and the boundary entropy value of the horizon i.e., $S = S_f + S_h$. The laws of thermodynamics suggests 
\begin{eqnarray}
&& \frac{dS}{dn}\geq 0, \\&&
\frac{d^2S}{dn^2} < 0, \label{therm}
\end{eqnarray}
where $n=\ln{a}$. It is straightforward to prove that \citep{jamil, pan}

\begin{equation}\label{Sdn}
S_{,n} \propto \frac{(H_{,n})^2}{H^4}.
\end{equation}
The right hand side is a full square and is always positive. Differentiating once more Eq. (\ref{Sdn}) produces 

\begin{equation}\label{Entropy_Psi}
S_{,nn} = 2S_{,n}\left(\frac{H_{,nn}}{H_{,n}}-\frac{2H_{,n}}{H}\right) = 2S_{,n}\Psi.
\end{equation}

\begin{figure}
\begin{center}
\includegraphics[width=0.40\textwidth]{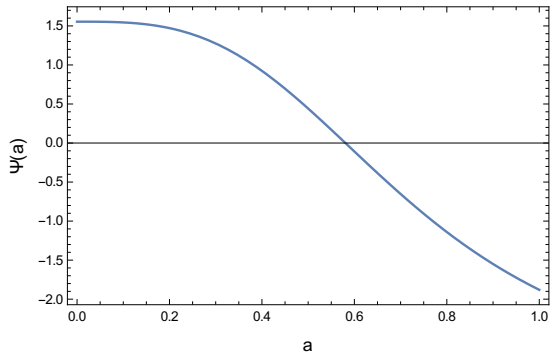}
\caption{Evolution of $\left(\frac{H_{,nn}}{H_{,n}}-\frac{2H_{,n}}{H}\right)=\Psi$ as a function of $a$ for the best fit parameter values $h_{0} \sim 0.71$, $C_{1} \sim 2.913$ and $\delta \sim 0.67$.}
\label{psiplot}
\end{center}
\end{figure}

From Eq. (\ref{Entropy_Psi}) we note that a thermodynamic equilibrium ($S_{,nn} < 0$) requires $\Psi$ to be negative. We plot the evolution of $\Psi$ for the reconstructed kinematic model as a function of $a$ in Fig. \ref{psiplot}. $\Psi$ shows an evolution in the negative domain during the late time evolution of the universe. The evolution follows closely a $\Lambda$CDM behavior including a smooth transition from positive into negative values. However, the epoch of transition around which the universe has moved into a thermodynamic equilibrium is expected to be parameter sensitive. This behavior as well as any of the discussed reconstructed evolution are applicable to any other extended theory of gravity as well, albeit with relevant constraints. This is because the entire scheme of the parametrization depends on a kinematic quanitity, the statefinder. For instance, the thermodynamic behavior and equilibrium discussed above is also discussed in comprehensive details quite recently, in the context of the extended BD theories that mimic the runnng vacuum behavior\citep{solathermo}.

\section{Brans-Dicke-Higgs : Reconstruction of Scalar Field Profiles and their Interactions}
In the previous section we discussed the significance and viability of cosmological reconstruction from
a purely kinematic quantity, the statefinder parameter in this case. The reverse engineered evolution of the universe also allows a smooth transition from a decelerated into an accelerated expansion in the recent past. Now we study the field equations of the modified Brans-Dicke-Higgs setup and reconstruct the allowed lagrangian of the theory that can support such an evolution. On that note, we first write the Hubble and it's time derivative as a function of scale factor $a$ for a chosen set of parameter values very close to the best fit.

\begin{eqnarray}\label{hubblebd}
&& H(a) = a^{\frac{3}{20}}(3+a^{-3.06})^{\frac{1}{2}}, \\&&\nonumber
\dot{H} = (3+a^{-3.06})^{\frac{1}{2}} \Big\lbrace (3+a^{-3.06})^{\frac{1}{2}} \\&&
- \frac{3}{2}a^{-3.06}(3+a^{-3.06})^{-\frac{1}{2}} \Big\rbrace - a^{\frac{3}{10}}(3+a^{-3.06}). 
\end{eqnarray}

We assume that $\frac{\partial U}{\partial \phi} \sim \phi_{0}$ and $\frac{\partial U}{\partial \psi} \sim \psi_{0}$, where $\phi_0$ and $\psi_0$ are small constant parameters. In other words, the interaction term $U(\phi, \psi)$ is a slowly varying function of $\phi$ and $\psi$. Apart from the ansatz over the statefinder parameter, this is the only assumption we are postulating at the outset in order to study the system of equations. This accounts for additional restrictions over the interaction functional as can be understood by writing $U$ as a function of redshift $z$. The first derivative is
\begin{equation}
\frac{dU}{dz} = \frac{\partial U}{\partial \phi} \frac{d\phi}{dz} + \frac{\partial U}{\partial \psi}\frac{d\psi}{dz}.
\end{equation}
Even if $U(\phi, \psi)$ is a slowly varying function of $\phi$ and $\psi$, the strength of the interaction as a function of redshift also depends on the pair $\frac{d\phi}{dz}$ and $\frac{d\psi}{dz}$, whose evolution is also our subject of interest. Overall this indeed is a special case, but without this simplification directly solving the scalar evolution equations, analytically or numerically becomes non-trivial. Nevertheless, the assumption does not inspire any unphysicality as we demonstrate with an example. As discussed earlier, the cross-interaction $U(\phi, \psi)$ is inspired from an idea of interacting Dark Energy-Dark Matter scenario. There are more than a few theoretical models of the same in literature, some of which are further supported by phenomenological comparisons \citep{interdmde1, interdmde2, interdmde3}. One among the most popular models is characterized by an interaction potential of the form \citep{bertolamiinter}
\begin{equation}\label{crosspot1}
U(\psi,\phi) = e^{-\lambda \psi}P(\psi,\phi)
\end{equation} 

In general a suitable $P(\psi,\phi)$ is chosen such that phenomenological requirements are met. For instance, it can be proved \citep{bertolamiinter} that the features of the present universe is reproduced upto a fair approximation if $P(\psi,\phi)$ is a polynomial in $\psi$. In the present approach, using this form as an example we show that a slowly varying $U(\phi, \psi)$ simply allows us to constrain the otherwise free function $P(\psi,\phi)$ such that the reconstructed evolution fits in with observations perfectly. Using the above ansatz it is straightforward to show that

\begin{eqnarray}
&& \frac{\partial P}{\partial \psi} - \lambda P = \psi_{0} e^{\lambda \psi}, \\&&
\frac{\partial P}{\partial \phi} = \phi_{0} e^{\lambda \phi}.
\end{eqnarray}
This leads us to the conclusion that for the present setup
\begin{equation}\label{crosspot2}
P(\psi,\phi) = \frac{\phi_{0}}{\lambda}e^{\lambda\phi} + Q(\psi).
\end{equation} 

Using Eq. (\ref{hubblebd}) we first transform the time derivatives in the field Eqs. (\ref{eq:EoM-metric}), (\ref{eq:EoM-phi}) and (\ref{eq:EoM-psi}) into derivatves with respect to $x = \frac{1}{a}$ and then into derivatives with respect to redshift $z = x - 1$. A derivative with respect to $x$ is written as prime and a derivative with respect to redshift is written as overhead circle `$\circ$'. For the sake of brevity we demonstrate the coordinate transformations for $\psi$ only.

\begin{eqnarray}\label{psibd}
&&\dot{\psi} = \psi' a^{\frac{23}{20}}(3+a^{-3.06})^{\frac{1}{2}}, \\&&\nonumber
\ddot{\psi} = \psi' (3 + a^{-3.06})^{\frac{1}{2}} a^{\frac{23}{20}} \Big\lbrace (3+a^{-3.06})^{\frac{1}{2}} \\&&
- \frac{3}{2}a^{-3.06}(3+a^{-3.06})^{-\frac{1}{2}} \Big\rbrace - \psi'' a^{\frac{23}{10}}(3+a^{-3.06}), \\&&
\psi' = -\frac{\psi^{\circ}}{a^2},\\&&
\psi'' = \frac{\psi^{\circ\circ}}{a^4} + \frac{2\psi^{\circ}}{a^3}.
\end{eqnarray} 
Using Eq. (\ref{psibd}) we can write the evolution equation of the BD scalar field $\psi$ Eq. (\ref{eq:EoM-psi}) as a function of redshift

\begin{eqnarray}\label{psibda}\nonumber
&& \psi^{\circ\circ} = \frac{{\psi^{\circ}}^2}{2\psi} + \frac{\psi^{\circ}}{a^2}\Big[(3+a^{-3.06})^{-\frac{1}{2}} a^{-\frac{23}{20}} \Big\lbrace (3+a^{-3.06})^{\frac{1}{2}} \\&&\nonumber
- \frac{3}{2}a^{-3.06}(3+a^{-3.06})^{-\frac{1}{2}} \Big\rbrace + \frac{1}{a} \Big] + \frac{\psi_{0}\psi}{\omega}a^{\frac{17}{10}}\\&&\nonumber
(3+a^{-3.06})^{-1} + \frac{\psi}{\omega} \Big[3(3+a^{-3.06})^{-\frac{1}{2}} a^{\frac{17}{10}} \Big\lbrace (3+a^{-3.06})^{\frac{1}{2}} \\&&
- \frac{3}{2}a^{-3.06}(3+a^{-3.06})^{-\frac{1}{2}} \Big\rbrace - 3a^2 \Big].
\end{eqnarray}

Using $a = \frac{1}{1+z}$ we numerically solve Eq. (\ref{psibda}) as a function of redshift for the parameter choice of $\psi_{0} \sim O(10^{-1})$ and $\omega \sim 10^5$. Since BD theory is a varying $G$ theory, the rate of this variation can be compared with the observational limit and the resulting requiremment inspires the choice of initial conditions for a numerical solution of the geometric scalar field. Keeping in mind that $\frac{1}{\psi}$ behaves as an effective $G$ \citep{weinbook, pavon, sudiptadi}, we take the initial conditions to solve Eq. (\ref{psibda}) as
\begin{eqnarray}
&&\frac{\dot{G}}{G} = -\frac{\dot{\psi}}{\psi} = +\frac{k}{H} \mbox{where, $k \leq 1$ } \\&&
\Bigg\vert \frac{\dot{G}}{G} \Bigg\vert_{z=0} \equiv \Bigg\vert \frac{\dot{\psi}}{\psi} \Bigg\vert_{z=0} = \frac{k}{H_0} \leq 10^{-10} \mbox{per year.}
\end{eqnarray}

\begin{figure}
\begin{center}
\includegraphics[width=0.40\textwidth]{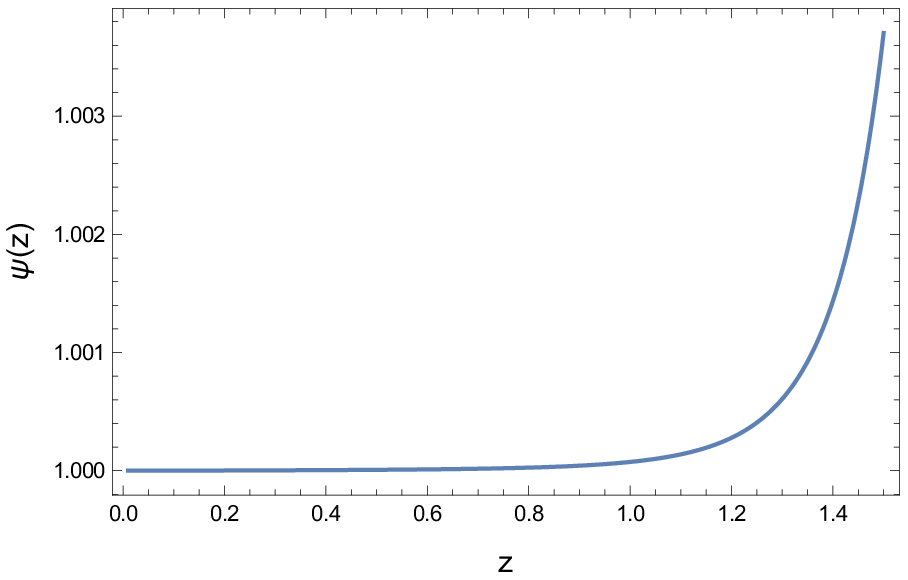}
\includegraphics[width=0.40\textwidth]{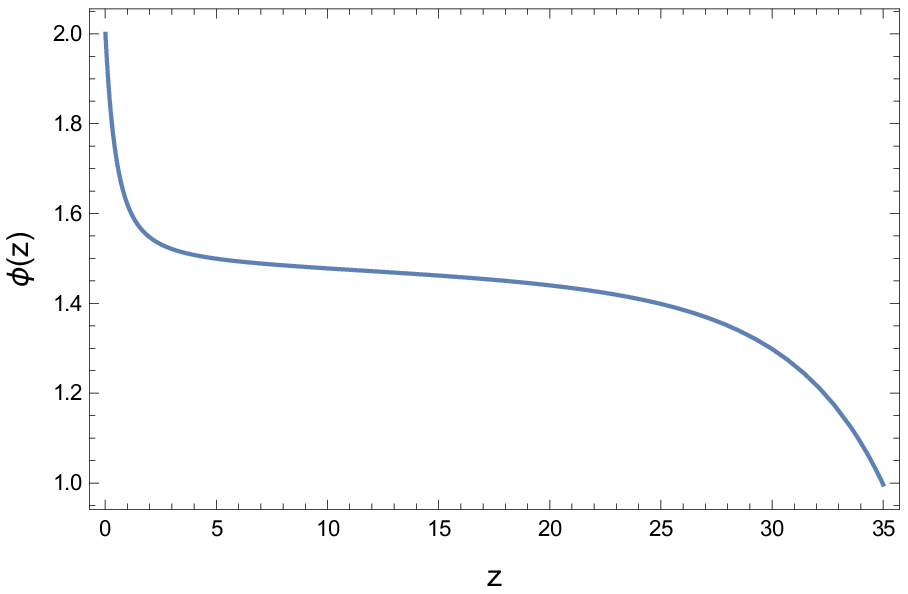}
\caption{Evolution of the Brans-Dicke Scalar Field $\psi(z)$ (Top Panel) and the Higgs Scalar Field $\phi(z)$ (Bottom Panel) for the best fit parameter values $h_{0} \sim 0.71$, $C_{1} \sim 2.913$ and $\delta \sim 0.67$.}
\label{scalars}
\end{center}
\end{figure}

We plot the evolution of $\psi$ in the top panel of Fig. \ref{scalars}. In a similar manner the Higgs scalar field evolution Eq. (\ref{eq:EoM-phi}) can be numerically solved as a function of redshift. The evolution is shown in the bottom panel of Fig. \ref{scalars}. It is evident from the graphs that at the present epoch of accelerated expansion, i.e., $(z \sim 0)$, the geometric {\it a.k.a.} Brans-Dicke scalar field $\psi$ is subdued compared to the scalar field that has a generalized derivative coupling with gravity or the Higgs field $\phi$. We can generalize the statement and note that in a two-scalar lagrangian theory of gravity similar to Eq. (\ref{eq:SMBDaction}) a late-time cosmic acceleration is generated by the scalar field having a derivative interaction. This claim is extended by the plot of $\phi(z)$ as it can be seen that for high redshift values the scalar also shows a sharply increasing behavior, very similar to the behavior around $z \sim 0$. For $z > 1$, $\phi$ behaves as a very slowly varying function of $z$ which is expected to be the epoch of a cosmic deceleration. During this epoch the BD scalar $\psi$ becomes dominant and exhibits a sharply increasing profile, for $z \sim 1$ and higher. As already discussed in the previous section, the evolution of the deceleration parameter suggests that $z_{t} \sim 1$ is the transition redshift around which the universe is expected to have a smooth transition from a decelerated into an accelerated expansion.  \\

We discuss that the interaction between $\phi$ and $\psi$ has an intriguing role in subduing one scalar field during acceleration and another during deceleration. This is realized by solving for the interaction $U (\phi,\psi)$ as a function of redshift using Eq. (\ref{eq:EoM-metric}). To do this we specify the Higgs scalar interaction potential at the outset which has a generic form
\begin{equation}\label{eq:VHiggs}
V = \frac{\mu^{2}}{2}\phi^{2} + \frac{\lambda}{4}\phi^{4},
\end{equation}
$\lambda$ is dimensionless while the dimension of $\mu^2 < 0$ is mass squared. Constraints on these parameters can be determined in a self-consistent manner in terms of the interaction coefficients in an extended BD lagrangian, using a power-law cosmological solution and $U \sim \psi \phi^{2}$. However in the present case, understanding of the behavior of the interaction requires a different approach. To this end we choose the Higgs potential at the outset where the vacuum expectation value (VEV) of the Higgs field $\phi$ can be written as
\begin{equation}\label{eq:HiggsMass}
M_H^2 = \left.\frac{\partial^2 V}{\partial\phi^2}\right|_{\phi=v} = \mu^2 + 3\lambda\,v^2 = -2\mu^2>0\,,
\end{equation}
defined at the minima of the potential $v = \sqrt{-\mu^2/\lambda}$. It is also insightful to mention that this VEV can be compared with the energy scale of $\Lambda$CDM. As meticulously proved \citep{sola}, an extended Brans Dicke theory can allow a Higgs field in the lagrangian and simulteneously satisfy the requirements of Electroweak theory only if the coefficient of the cross-interaction is chosen as a very small parameter. Before the discovery of Higgs, $\mu$ and $\lambda$ could be treated as free parameters to serve a theorist's pleasure. However, recent discoveries in this regard dictates that the VEV must be fixed such that experimental bounds on $W-$boson mass is satisfied, requiring
\begin{equation}
M_{\small W} \propto v,
\end{equation}
with $v \sim 246$ GeV. In the present case, we write the Higgs potential as 
\begin{equation}\label{higgsnontrivial}
V(\phi) = V_{0} - \frac{1}{2}m^{2}\phi^{2} + \frac{m^2}{24f^2}\phi^{4}.
\end{equation}
This is a little non-trivial form, as in we have chosen the coefficient of $\phi^2$ to be negative at the outset and therefore in this setup, $m^{2} > 0$. The VEV becomes $12f^2 \sim M_{W}^{2} \sim 6 \times 10^{4} GeV^{2}$ and one needs to fix the parameters of the theory accordingly. The dimesionless parameters of the theory are chosen as 
\begin{eqnarray}\nonumber
&& \theta = 2 \;\;;\;\; \varsigma = 1 \;\;;\;\; \xi = \frac{1}{6}.  
\end{eqnarray}
As a function of scale factor and redshift the interaction $U (\phi, \psi)$ can be written as

\begin{eqnarray}\label{interactionbd}\nonumber
&& U(\phi, \psi) = -3a^{\frac{3}{10}} (3 + a^{-3.06})\psi + 3 a^{-\frac{7}{10}}(3 + a^{-3.06})\psi^{\circ} \\&&\nonumber
+ \frac{\omega}{2}\frac{{\psi^{\circ}}^2}{\psi}a^{-\frac{17}{10}}(3 + a^{-3.06}) + \frac{1}{2}{\phi^{\circ}}^{2} a^{-\frac{17}{10}}(3 + a^{-3.06}) + V_{0} \\&&\nonumber
- \frac{1}{2}m^{2}\phi^{2} + \frac{m^2}{24f^2}\phi^{4} - 6\xi a^{\frac{3}{10}}(3 + a^{-3.06})\phi^{2} + 12\xi a^{\frac{3}{10}} \\&&\nonumber
(3 + a^{-3.06})\phi^{\circ}\phi + 9\theta a^{-\frac{7}{5}}(3 + a^{-3.06})^{2} \frac{{\phi^{\circ}}^{2}}{\phi^{2}} + 6(\theta - \varsigma) \\&&\nonumber
\Big[(3 + a^{-3.06})^{\frac{1}{2}} \Big\lbrace(3 + a^{-3.06})^{\frac{1}{2}} - \frac{3}{2}a^{-3.06}(3 + a^{-3.06})^{-\frac{1}{2}} \Big\rbrace \\&&\nonumber
- a^{\frac{3}{10}}(3 + a^{-3.06})\Big] \frac{{\phi^{\circ}}^2}{\phi^{2}}a^{-\frac{17}{10}}(3 + a^{-3.06}) + 6(\theta - \varsigma) \\&&\nonumber
a^{\frac{13}{10}}(3 + a^{-3.06})\Big[-\phi^{\circ} (3 + a^{-3.06})^{\frac{1}{2}} a^{\frac{3}{20}} \Big\lbrace (3 + a^{-3.06})^{\frac{1}{2}} - \\&&\nonumber
\frac{3}{2} a^{-3.06} (3 + a^{-3.06})^{-\frac{1}{2}} \Big\rbrace + \Big(\frac{\phi^{\circ\circ}}{a^4} + \frac{2 \phi^{\circ}}{a^3} \Big) a^{\frac{23}{10}}(3 + a^{-3.06})\Big] \\&&
\frac{\phi^{\circ}}{a^{2}\phi^{2}} - 6(\theta - \varsigma)a^{-\frac{12}{5}}(3 + a^{-3.06})^{2} \frac{{\phi^{\circ}}^{3}}{\phi^{3}}.
\end{eqnarray}

\begin{figure}
\begin{center}
\includegraphics[width=0.40\textwidth]{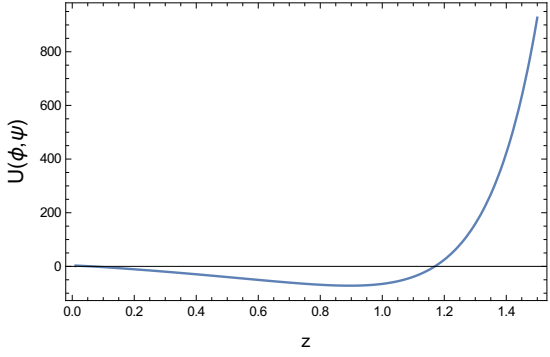}
\caption{Strength of the interaction between the Brans-Dicke Scalar Field $\psi(z)$ and the Higgs Scalar Field $\phi(z)$, $U(\psi, \phi)$ vs $z$, for the best fit parameter values $h_{0} \sim 0.71$, $C_{1} \sim 2.913$ and $\delta \sim 0.67$.}
\label{interaction}
\end{center}
\end{figure}
At this point it is important to recall that the BD scalar $\psi$ has a dimension of mass squared in natural units. The Higgs scalar $\phi$ has mass dimension $1$ in natural units. From the action in Eq. (\ref{eq:SMBDaction}), it can be seen that the dimensions of the scalars put some serious constraint over the functional form of the cross interaction $U(\phi, \psi)$. Given the dimension of $\sqrt{-g}d^{4}x$, all terms inside the square bracket including $U(\phi, \psi)$ must have a dimesion of mass$^4$. The most straightforward choice for the interaction term satisfying the dimensional requirement is $U(\phi, \psi) \propto \psi \phi^{2}$ which is considered in literature at a great length \citep{sola}. However, other forms can be viable as well, for instance, a from similar to Eqs. (\ref{crosspot1}) and (\ref{crosspot2}) provided the dimensional requirement is satisfied. However, we primarily focus on the evolution of this interaction as a function of redshift in this manuscript. \\
A numerical solution leads to the graph in Fig. \ref{interaction}, $U(\phi, \psi)$ vs $z$. We take the parameter $V_{0} \sim O(1)$. It appears from the graph that a strong interaction between the two fields dominates the era of cosmic deceleration interaction and falls off sharply from a large value around $z_{t} \sim 1$, the transition redshift, when the epoch of acceleration is supposed to switch itself on. In other words, $U$ effectively acts as a switch of smooth transition between deceleration and acceleration. The limitation, however, is that the discussed evolution of $U$ as a function of redshift is only valid upto a particular value of redshift, $z \sim 2.5$. Since we compare the theoretical model with observatinal data at low redshift only, a constant or a near constant statefinder parameter can be a parametrization of the late time acceleration and the earlier deceleration only upto a certain extent. Therefore it is not enough to comment on the behavior of $U$ in greater past, i.e., for very high value of redshift or around $a \sim 0$. Investigation of an unified evolution of the Universe and the role of the interaction therein will form a more novel discussion and requires a better parametrization of the statefinder as a function of $z$, instead of just a constant. \\

In a nutshell, we note that during the late-time evolution of the universe, due to the dominance of the Higgs field and it's interaction, the structure of the theory effectively becomes similar to that of a non-minimally coupled self-interacting scalar field theory. The non-minimally coupled theory interacts with the lagrangian derivatively as well. The BD scalar, it's kinetic term and the cross-interaction term are subdued and one can ignore $\frac{1}{2}R\psi - \frac{\omega}{2\psi}g^{\mu\nu}\partial_{\nu}\psi\partial_{\mu}\psi + U(\phi , \psi)$ compared to the rest of the action, given by

\begin{eqnarray}\nonumber
&& S_{acc} \sim \int d^{4}x\sqrt{-g}\Bigg[\xi R\phi^{2} - \frac{1}{2} g^{\mu\nu}\partial_{\mu}\phi\,\partial_{\nu}\phi \\&&
+ \frac{1}{\phi^{2}}S_{\mu\nu}\partial^{\mu}\phi\partial^{\nu}\phi - V_{0} + \frac{1}{2}m^{2}\phi^{2} - \frac{m^2}{24f^2}\phi^{4}\Bigg].
\end{eqnarray}

On the other hand, during a decelerated expansion, the BD scalar takes the dominant role compared to the Higgs scalar. In such a case, the theory essentially becomes a standard Brans-Dicke theory, alongwith a dominant interaction potential $U (\phi, \psi)$.

\begin{equation}
S_{dec} \sim \int d^{4}x\sqrt{-g}\Bigg[\frac{1}{2}R\psi - \frac{\omega}{2\psi}g^{\mu\nu}\partial_{\nu}\psi\partial_{\mu}\psi + U(\phi , \psi)\Bigg].
\end{equation}

Since we were interested in a cosmological setup where scalar fields dominate over all the other possible field sources, we included no additional fluid energy momentum source in the action. It is true that for a correct description of the present accelerating universe one must use the scalar fields, pressureless dust matter as well as some form of Dark Energy. However, as discussed earlier, the present work is an adventurous attempt, if only mathematical, to write both the required Dark Matter and Dark Energy sectors using just the scalars and their interactions. Due to the departure of the theory from standard GR, the modified field equations in Eq. (\ref{eq:EoM-metric}) and Eq. (\ref{eq:EoM-psi}) can be rewritten as effective density and pressure

\begin{eqnarray}\nonumber
&& \rho_{eff} = -\frac{1}{\psi} \Bigg[{{3}H{\dot{\psi}}}-\frac{\omega}{2}\frac{\dot{\psi}^{2}}{\psi} + U(\phi, \psi) -\frac{1}{2}\dot{\phi}^{2}-V(\phi) \\&&\nonumber
+ 6\xi H^{2}\phi^{2} + 12\xi H \dot{\phi}\phi - {9}\theta H^{2}\frac{\dot{\phi}^{2}}{\phi^{2}} -  6(\theta-\varsigma)\dot{H}\frac{\dot{\phi}^{2}}{\phi^{2}} \\&&
+ 6(\theta-\varsigma)H\frac{\dot{\phi}\ddot{\phi}}{\phi^{2}} - 6(\theta -\varsigma){H}\frac{\dot{\phi}^{3}}{\phi^{3}}\Bigg],
 \label{effdens}
\end{eqnarray}
and
\begin{equation}
p_{eff} = \frac{2}{3} \Bigg[6{H}^{2} - \omega \frac{\ddot{\psi}}{\psi}+\frac{ \omega}{2} \frac{\dot{\psi}^{2}}{{\psi}^{2}}-3H\omega\frac{\dot{\psi}}{\psi}+ \frac{\partial U}{\partial \psi}\Bigg] - \rho_{eff}. \label{effpress}
\end{equation}

\begin{figure}
\begin{center}
\includegraphics[width=0.40\textwidth]{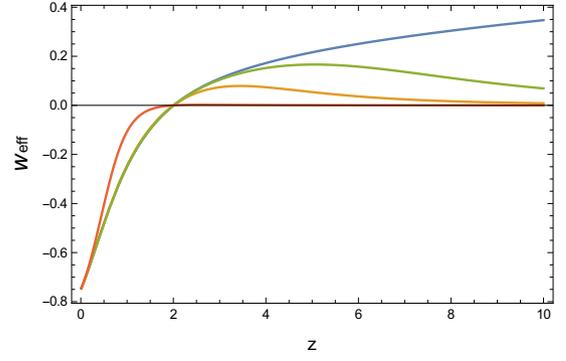}
\caption{Evolution of effective equation of state parameter as a function of $z$, for the best fit parameter values $h_{0} \sim 0.71$, $C_{1} \sim 2.913$ and $\delta \sim 0.67$, using Eqs. (\ref{effdens}) and (\ref{effpress}). Blue curve is for $\omega = 50000$, Green curve is for $\omega = 60000$, Orange curve is for $\omega = 70000$ and Red curve is for $\omega = 100000$.}
\label{EOSBD}
\end{center}
\end{figure}

In Fig. \ref{EOSBD} we plot the effective equation of state $w_{eff} \sim \frac{p_{eff}}{\rho_{eff}}$ as a function of redshift. The plots indicate that at low redshift there is a repulsive effect due to negative pressure leading to the present accelerated expansion. $w_{eff}$ reaches a value $\sim -0.8$ around the current epoch $z \sim 0$. The behavior of $w_{eff}$ for $z \geq 2$ is however, sensitive over the choice of the Brans-Dicke parameter $\omega$. For $\omega = 50000$, the evolution is shown by the blue curve and it suggests that $w_{eff}$ rolls into a positive domain in high redshift region, which is suggestive of a radiation dominated epoch. However, as one slightly increases the value of $\omega$, (Green curve is for $\omega = 60000$, Orange curve is for $\omega = 70000$ and Red curve is for $\omega = 100000$) $w_{eff}$ at a higher redshift moves closer to a zero value which suggests a dust dominated era immediately prior to the accelerated expansion. Effectively this indicates that the higher value of $\omega$ we choose, the closer the theory is to General Relativity and more similar the solution behaves as compared to a $\Lambda$CDM cosmology. \\

Next, we focus on the idea of structure formation in the BD-Higgs setup. Evolution of perturbation and the growth of matter overdensity in the context of generalized BD theory is an intriguing subject and has received rigorous attention quite recently. We review the basic equations here in a nutshell for the sake of a compact idea. One usually works with an FRW metric written as 
\begin{equation}
ds^2 = a^2(\tau)[-d\tau^2+(\delta_{ij}+h_{ij})dx^idx^j],
\end{equation}
$\tau$ being the conformal time and $h_{ij}$, a perturbation on the spatial part of the metric defined as,
\begin{equation}\label{eq:MainhFourier}
h_{ij}(\tau,\vec{x}) = \int d^3k\, e^{-i\vec{k}\cdot\vec{x}}\left[\hat{k}_i\hat{k}_j h(\tau,\vec{k})+\left(\hat{k}_i\hat{k}_j-\frac{\delta_{ij}}{3}\right)6\xi(\tau,\vec {k})\right].
\end{equation}
The perturbation is defined in momentum space. The trace part is defined as $h\equiv \delta^{ij}h_{ij}$ whereas the traceless part is written by $\xi$. One defines $\mathcal{H}\equiv a^\prime/a$ where a prime denotes derivative with respect to the conformal time. Under the assumption of $k^2\gg \mathcal{H}^2$, (for a detailed analysis, we refer to the monograph by \cite{sola4}) one can derive the perturbation equations
\begin{equation}\label{eq:Mainsimpli0}
\delta_{m}' = -\frac{h'}{2}\,.
\end{equation}
\begin{equation}\label{eq:Mainsimpli1}
k^2\delta \psi+\frac{h'}{2}{\psi}^\prime =\frac{8 \pi G_N}{3+2\omega_{BD}}a^2 {\rho}_m\delta_m\,,
\end{equation}
\begin{equation}\label{eq:Mainsimpli2}
{\psi}(\mathcal{H}h' - 2\xi k^2) + k^2\delta\psi + \frac{h'}{2}{\psi}' = 8\pi G_N a^2 {\rho}_m\delta_m\,,
\end{equation}
\begin{equation}\label{eq:Mainsimpli3}
2k^2\delta \psi + {\psi}'h' + {\psi}\left(h'' + 2h' \mathcal{H} - 2k^2 \xi\right)=0\,.
\end{equation}
A combination of these equations leads one to a generalized evolution of the growth of matter overdensity contrast at the linear level
\begin{equation}\label{eq:ExactPerturConfTime}
\delta_{m}'' + \mathcal{H}\delta_{m}' - \frac{4\pi G_N a^2}{\psi}\rho_{m}\delta_{m} \left(\frac{4+2\omega_{BD}}{3+2\omega_{BD}}\right)=0\,.
\end{equation}
or
\begin{equation}\label{eq:ExactPerturConfTime2}
\delta_{m}'' + \mathcal{H}\delta_{m}' - 4\pi G_{eff}(\bar{\psi}) a^2\,\bar{\rho}_m\delta_m=0\,.
\end{equation}

We find it convenient to write the matter overdensity as a function of scale factor $a$ and write Eq. (\ref{eq:ExactPerturConfTime2}) as
\begin{equation}\label{eq:ExactPerturscale}
\delta_{m}^{\circ\circ} + \delta_{m}^{\circ}\left(\frac{\ddot{a}}{\dot{a}^2}+\frac{2}{a}\right) - 4\pi \frac{G_{eff}(\psi)}{\dot{a}^2}\rho_{eff}\delta_m = 0\,.
\end{equation}

\begin{figure}
\begin{center}
\includegraphics[width=0.40\textwidth]{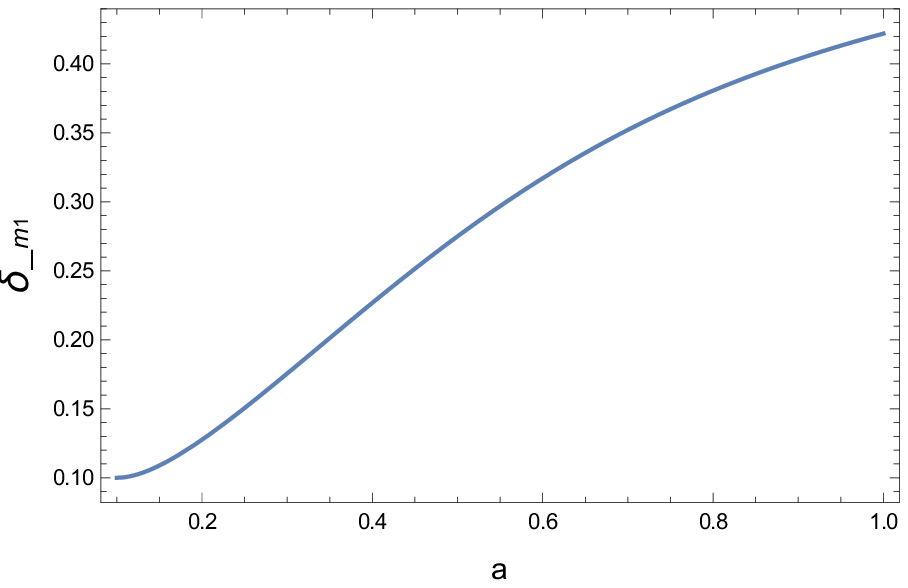}
\includegraphics[width=0.40\textwidth]{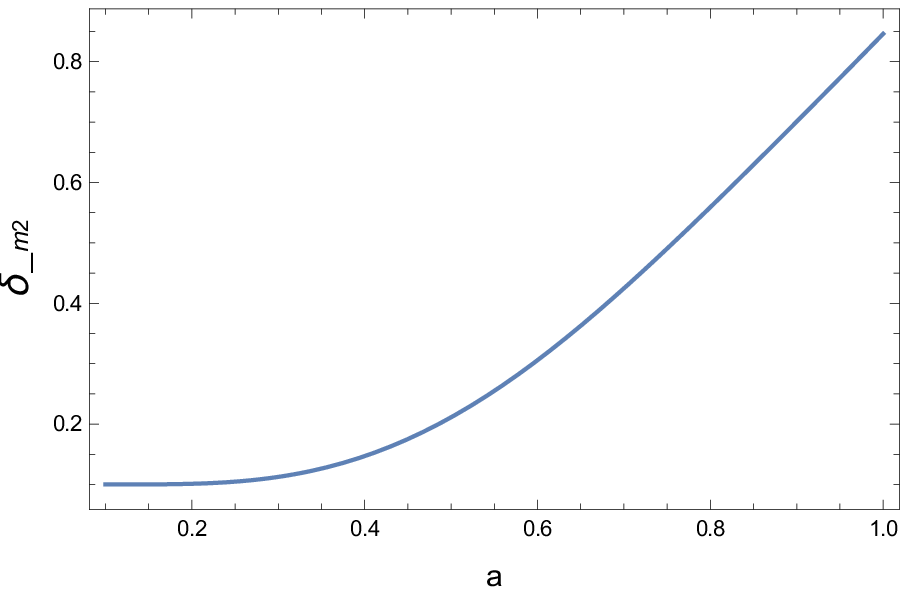}
\caption{Evolution of growth of matter over-density $\delta_m$ as a function of scale factor for the best fit parameter values $h_{0} \sim 0.71$, $C_{1} \sim 2.913$ and $\delta \sim 0.67$ with the reconstructed density profile as in Eq. (\ref{effdens}). $\delta_{m1}$ vs $a$ is a plot for $z \sim 0$, where $G$ can be approximated as a constant. $\delta_{m2}$ vs $a$ is a plot for higher $z$, where $G \sim \frac{1}{\psi}$.}
\label{overdensityBD}
\end{center}
\end{figure}

An overhead $\circ$ denotes derivative with respect to scale factor $a$ and {\it dot} is derivative with respect to cosmic time $t$. We use the connection between cosmic time and conformal time $\frac{dt}{d\tau} = a$. Therefore, $\mathcal{H} = \dot{a}$. We also write $\frac{4\pi G_N}{\psi}\left(\frac{4+2\omega_{BD}}{3+2\omega_{BD}}\right) = G_{eff}$. The somewhat non-trivial notion comes in the form of a $\rho_{eff}$ in Eq. (\ref{eq:ExactPerturscale}), since there is no direct fluid matter included in the action. However, we emphasize once more that our aim is to use the scalar fields in a manner that they can suffice for an effective fluid description in the field equation. This is seen from the field Eq. (\ref{eq:EoM-metric}) by keeping $3H^2$ on the LHS and treating the rest of the term on RHS as an effective fluid contribution to the energy density. This is actually not too unphysical since, a Brans-Dicke theory or some it's generalizations are known to generate accelerating solutions even without matter. Moroever, in a completely general context, it can be proved that a scalar field with suitable choice of self-interactions can mimic the evolution of many a kind of realistic matter distribution \citep{goncamoss}. The matter over-density $\delta_m$ at the linear level as a function of cosmic time therefore can be approximately written by the differential equation
\begin{equation}\label{delm_eqBD}
\ddot{\delta}_m + 2H\dot{\delta}_m = 4\pi G_{eff}(\psi) \rho_{eff}\delta_m,
\end{equation}
with $\frac{4\pi G_N}{\psi}\left(\frac{4+2\omega_{BD}}{3+2\omega_{BD}}\right) = G_{eff}$. Eq. (\ref{delm_eqBD}) is quite similar to the standard evolution Eq. (\ref{delm_eq}), however, with expected generalizations in $\rho_{eff}$ and a varying $G_{eff}$. Using the numerical solution found for $\psi$, we study $\delta_m$ numerically, as a function of scale factor $a$. We solve the equation with initial conditions $a_i = 0.01$ and the initial values are fixed as, $\delta_m(a_i) = 0.01$ and $\dot{\delta}_m(a_i) = 0$. The evolution of $\delta_m$ vs $a$ for the best fit parameter values $h_{0} \sim 0.71$, $C_{1} \sim 2.913$ and $\delta \sim 0.67$ are studied for two different cases, and is shown in Fig. \ref{overdensityBD}. $\delta_{m1}$ vs $a$ shows a plot of the matter overdensity for late-times, $z \sim 0$, around which $\psi$ behaves almost as a constant and $G$ can be thought of as a constant as well. $\delta_{m2}$ vs $a$ shows the plot of the matter overdensity when the BD scalar field is the dominating scalar in the theory and is given by the solution of the Eq. (\ref{psibda}). $G(\psi)$ is then given by Eq. (\ref{eq:Geff}). We note that even for the best fit parameter values, there are some departure of the evolution from a $\Lambda$CDM cosmology.

\begin{figure}
\begin{center}
\includegraphics[width=0.40\textwidth]{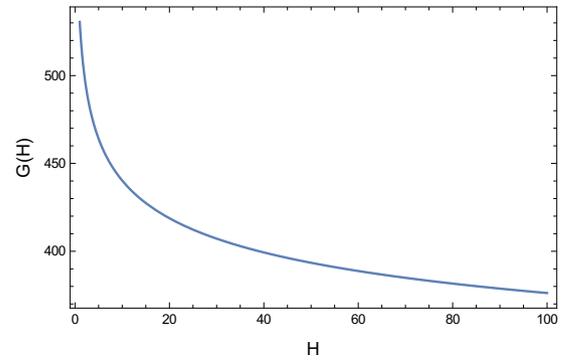}
\caption{Evolution of the gravitational constant as a function of Hubble for the best fit parameter values $h_{0} \sim 0.71$, $C_{1} \sim 2.913$ and $\delta \sim 0.67$.}
\label{G}
\end{center}
\end{figure}

From Eq. (\ref{eq:Geff}) we plot the effective gravitational constant $G_{eff}$ as a function of Hubble, in Fig. \ref{G}. The graph clearly suggests that $G(H)$ is a decreasing function of $H$. In other words, gravitational interaction is stronger in earlier epoch where Hubble rate is low. With an accelerated expansion Hubble increases and the interaction weakens, making the extended Brans-Dicke-Higgs theory considered here an asymptotically free theory. The evolution predicts similar physics when compared to the models of gravitational coupling with dynamical vacuum \citep{solag1, solag2}. The Brans-Dicke scalar field in the models of dynamic vacuum is predicted to be a slowly varying function of time which is consistent in the present model as well, atleast in the late-time epoch.

\section{Discussion and conclusion}\label{s3}
The present work explores cosmology in a generalized Brans-Dicke theory. In some sense the work has a motivation to glorify some aspects of scalar-tensor generalizations using essence from the Standard Model of particle physics, whose success relies largely on the features supporting a Higgs interaction potential of the constituent field. However, despite numerous attempts, genesis and structure of the Standard model in the presence of a gravitational interaction remains an unresolved problem. In addition to this, one must keep in mind that regular issues of modern cosmology such as the Cosmological Constant problem suggests that our understanding of the cosmic evolution remains incomplete. It is therefore a natural curiosity to consider an effective theory which can exhibit basic ensembles of particle physics as well as gravity. \\

A form of this extended theory was originally considered \citep{sola} in an attempt to interpret the Higgs interaction as an artefact of scalar fields and power-law cosmological solutions in Brans-Dicke theory. The so-called extended Brans-Dicke-Higgs action has the geometric BD-field interacting with a Higgs scalar field. The Higgs too interacts with the lagrangian derivatively and non-derivatively. We do not specify the interaction profiles of the Brans-Dicke scalar or the Higgs Scalar, or their mutual interaction at the outset. We work out a simple method of cosmological reconstruction from a kinematic quantity, the statefinder parameter, a dimensionless combination of the scale factor and it's higher order derivatives. The reconstruction leads us to write the Hubble function as a function of scale factor, or equivalently as a function of cosmic redshift. Thereafter we use the solution for Hubble to study the evolution of the coupling functions of the scalar fields in the lagrangian of the theory. Essentially, this exercise gives us a glimpse of the interacton profiles of the BD scalar and the Higgs scalar in the low-redshift regime; i.e., in the present epoch of accelerated expansion and the epoch immediatey prior to it. \\

The reconstruction produces cosmological solutions consistent with observational data as we confirm by estimating the model parameters from observational dataand studying the confidence contours in comparison with the best fit values in the parameter space. The data sets comes from the Supernova distance modulus-luminosity distance measurement, the Hubble parameter measurements (OHD) and the Baryon Acoustic Oscillation (BAO) data. We first note that the evolution of the deceleration parameter for the reconstructed model suggests that $z_{t} \sim 1$ is the transition redshift around which the universe goes through a smooth transition from a decelerated into an accelerated expansion. This is consistent with astrophysical observations. The value of the deceleration parameter ($\sim -0.62$ around  $z = 0$) agrees very well with present data, however, the evolution of jerk is non-trivial and signals a clear departure from a $\Lambda$CDM model. The statefinder parameter is estimated to be very close to zero. The BD scalar $\psi$ is found to be subdued for low redshift $z < 1$ and dominant for $z > 1$. On the other hand, the Higgs scalar $\phi$ is dominant in late-time era and for $z > 1$, $\phi$ behaves as a very slowly varying function of $z$. This can define a clear boundary between the epoch of a cosmic deceleration and acceleration. The deceleration is mainly driven by the BD scalar $\psi$ which becomes dominant for any value of $z \sim 1$ and higher. On the other hand around the redshift of transition the Higgs field $\phi$ starts dominating and the accelerated expansion is switched on. A strong interaction between the two fields dominates the era of cosmic deceleration only to fall off sharply during the era of late time acceleration. The interaction between the two scalar $\psi$ and $\phi$ therefore can resist a cosmic acceleration in a two-scalar lagrangian theory of gravity similar to the action in Eq. (\ref{eq:SMBDaction}). A late-time cosmic acceleration is driven by the scalar field having a derivative interaction in the lagrangian. Interesting probability can also be thought of regarding the cosmic acceleration in early universe from the evolution of $\phi(z)$, as for very high redshift values $\phi$ also shows a dominating behavior. As we have not assumed any fluid source of energy momentum at the outset, we define an effective density and pressure from the field equations of the theory. The evolution of the effective density and pressure and the effective equation of state clearly indicates that the present era of accelerated expansion of the universe is driven by a combination of fundamental matter fields which effectively generates a negative pressue. At a relatively higher redshift the universe approaches an era dominated by dust or radiation ($w_{eff} \geq 0$). We also note that at high redshift the behavior of $w_{eff}$ is sensitive over the choice of the Brans-Dicke parameter $\omega$, as in for a higher choice of $\omega$, the evolution of $w_{eff}$ follows a $\Lambda$CDM cosmological behavior more closely. Effectively this indicates the already established fact that the higher value of $\omega$ we choose, the closer the theory is to General Relativity. Evolution of perturbation and the growth of matter overdensity in the context of generalized BD theory is an intriguing subject and has received rigorous attention quite recently. We review the basic equations in this work and stress upon the fact that at the linear level of matter density perturbation, the evolution equation is quite similar to the standard evolution, with some interesting generalizations in effective density and a varying $G_{eff}$. The solution of the generalized equation shows some departure in the evolution of matter density contrast compared to a $\Lambda$CDM evolution. The reconstructed model in general obeys the requirement for a thermodynamic equilibrium of an expanding universe, irrespective of the theory under consideration, since the Hubble and scale factor are found straightaway from a kinematic approach. In this context a mathematical relation between the total entropy of the universe and the cosmological scale factor is written and the condition for thermodynamic equilibrium is studied. We note that the universe has encountered a smooth transition at some epoch of transition from a non-equilibrium into a thermodynamic equilibrium in recent past.  \\

This work demonstrates a simple approach to extract information from an otherwise complicated non-linear problem. The cosmological exact solution found from the kinematic approach is completely general and can be equally useful in any modified or generalized theory of gravity one is interested in. The variation of Newtonian Constant $G$ can probably lead one to consider time variation of `{\it residual vacuum dynamics} and in extension, a possibility that all the other fundamental constants are slowly varying with cosmic time as well. This is evocative of a vintage idea to tackle the `cosmological constant problem' \citep{bjorken}, and will be addressed in future works. \\

\section*{Acknowledgement}
The author thanks Prof. Koushik Dutta for insightful discussions and comments and Purba Mukherjee for fruitful discussions regarding the MCMC code. \\

{\bf Data Availability Statement} This manuscript has no associated data or the data will not be deposited. The datasets generated during and/or analyzed during the current study are available from the corresponding author on reasonable request.

\bibliographystyle{unsrt}

\begin{thebibliography}{}

\bibitem[\protect\citeauthoryear{Ade et. al.}{2014}]{planck}
Ade P. A. R. et. al., Planck collaboration, 2014, Astron. Astrophys. 571, A16.

\bibitem[\protect\citeauthoryear{Alam et. al.}{2003}]{alam}
Alam U., Sahni V., Saini T. and Starobinsky A. A., 2003, Mon. Not. Roy. Astron. Soc. 344 : 1057.

\bibitem[\protect\citeauthoryear{Alexander, Barrow and Magueijo}{2016}]{alexander1}
Alexander S., Barrow J. D. and Magueijo J., 2016, Class. Quant. Grav. 33, no.14, 14LT01.

\bibitem[\protect\citeauthoryear{Alexander, Marciano and Smolin}{2014}]{alexander2}
Alexander S., Marcian\`o A. and Smolin L., 2014, Phys. Rev. D. 89, 065017.

\bibitem[\protect\citeauthoryear{Amendola et. al.}{2012}]{amendola}
Amendola L., Leite  A., Martins C., Nunes  N., Pedrosa P. and Seganti A., 2012, Phys.
Rev. D 86, 063515.

\bibitem[\protect\citeauthoryear{Amendola}{1993}]{amendolan}
Amendola L., Phys. Lett. B, 1993, 301, 175.

\bibitem[\protect\citeauthoryear{Amendola}{1999}]{amendoo}
Amendola L., Phys.Rev. D. 1999, 60, 043501.

\bibitem[\protect\citeauthoryear{Anderson et al.}{2012}]{bossanderson}
Anderson L. et al. (BOSS collaboration), 2012, Mon. Not. Roy. Astron. Soc. 441, 24.

\bibitem[\protect\citeauthoryear{Aviles et. al.}{2012}]{aviles}
Aviles A., Gruber C., Luongo O. and Quevedo H., 2012, Phys. Rev. D 86, 123516.

\bibitem[\protect\citeauthoryear{Atkins and Calmet}{2013}]{atkins}
Atkins M. and Calmet X., Phys. Rev. Lett. 2013, 110, 051301 .

\bibitem[\protect\citeauthoryear{Bak and Rey}{2000}]{bak}
Bak D. and Rey S. J., 2000, Class. Quant. Grav. 17, L83.

\bibitem[\protect\citeauthoryear{Banerjee and Sen}{1997}]{nb1}
Banerjee N. and Sen S., 1997, Phys. Rev. D 56, 1334.

\bibitem[\protect\citeauthoryear{Banerjee and Pavon}{2001}]{nb2}
Banerjee N. and Pavon D., 2001, Class. Quant. Grav. 18, 593 ; Phys. Rev. D 63, 043504.

\bibitem[\protect\citeauthoryear{Banerjee and Pavon}{2001}]{pavon}
Banerjee N. and Pavon D., Phys. Rev. D. 63, 043504.

\bibitem[\protect\citeauthoryear{Barker}{1978}]{bark}
Barker B. M., 1978, Astrophys. J. 219, 5.

\bibitem[\protect\citeauthoryear{Bento et. al.}{2002}]{dmdeint3}
Bento M., Bertolami O. and Sen A., 2002, Phys. Rev. D66, 043507.

\bibitem[\protect\citeauthoryear{Bergmann}{1968}]{berg}
Bergmann P. G., 1968, Int. J. Theor. Phys. 1, 25.

\bibitem[\protect\citeauthoryear{Bernstein and Jain}{2004}]{berns}
Bernstein G. M. and Jain B., 2004, Astrophys. J. 600, 17.

\bibitem[\protect\citeauthoryear{Bertolami and Martins}{2000}]{bert}
Bertolami O. and Martins P., 2000, Phys. Rev. D. 61, 064007.

\bibitem[\protect\citeauthoryear{Bertolami et. al.}{2012}]{bertolamiinter}
Bertolami O., Carrilho P. and P\'aramos J, 2012, Phys. Rev. D 86, 103522.

\bibitem[\protect\citeauthoryear{Beutler et. al.}{2011}]{6dF}
Beutler F. et. al., 2011, Mon. Not. Roy. Astron. Soc. 416, 3017.

\bibitem[\protect\citeauthoryear{Bezeukov and Shaposhnikov}{2008}]{bez}
Bezrukov F. and Shaposhnikov M., 2008, Phys. Lett. B. 659, 703.

\bibitem[\protect\citeauthoryear{Betoule et al.}{2014}]{jla}
Betoule M. et al., 2014, Astron. Astrophys. 568, A22.

\bibitem[\protect\citeauthoryear{Bilic et. al.}{2002}]{dmdeint2}
Bilic N., Tupper G. B. and Viollier R. D., Phys. Lett. B535, 17.

\bibitem[\protect\citeauthoryear{Bjorken}{2001}]{bjorken}
Bjorken J. D., 2001, Phys. Rev. D. 64, 085008.

\bibitem[\protect\citeauthoryear{Blake et. al.}{2012}]{ohd5}
Blake C. et. al., 2012, Mon. Not. Roy. Astron. Soc. 425, 405.

\bibitem[\protect\citeauthoryear{Brans and Dicke}{1961}]{BD}
Brans C. H. and Dicke R. H., 1961, Phys. Rev. 124, 925.

\bibitem[\protect\citeauthoryear{Busti et. al.}{2015}]{busti}
Busti V. C., De la Cruz-Dombriz \'A., Dunsby P. K. S. and S\'aez-G\'omez D., 2015, Phys. Rev. D 92, 123512.

\bibitem[\protect\citeauthoryear{Capozziello and Lambiase}{1999}]{capopo}
Capozziello S. and Lambiase G., Gen. Rel. Grav. 1999, 31, 1005.

\bibitem[\protect\citeauthoryear{Cattoen and Visser}{2007}]{catt}
Cattoen C. and Visser M., 2007, Class. Quant. Grav. 24, 5985.

\bibitem[\protect\citeauthoryear{Chen et. al.}{2016}]{chen}
Chen Y., Ratra B., Biesiada M., Li S. and Zhu Z. H., 2016, Astrophys. J. 829, 61.

\bibitem[\protect\citeauthoryear{Chuang and Wang}{2013}]{ohd3}
Chuang C. H. and Wang Y., 2013, Mon. Not. Roy. Astron. Soc. 435, 255.

\bibitem[\protect\citeauthoryear{Clarkson and Zunckel}{2010}]{clarkson}
Clarkson C. and Zunckel C., 2010, Phys. Rev. Lett. 104, 211301.

\bibitem[\protect\citeauthoryear{Clifton and Barrow}{2006}]{clifton1}
Clifton T. and Barrow J. D., 2006, Phys. Rev. D 73, 104022.

\bibitem[\protect\citeauthoryear{Copeland, Sami and Tsujikawa}{2006}]{copeland}
Copeland E. J., Sami M. and Tsujikawa S., 2006, Int. J. Mod. Phys. D 15, 1753.

\bibitem[\protect\citeauthoryear{Crittenden et. al.}{2009}]{critt}
Crittenden R., Pogosian L. and Zhao G., 2009, JCAP 0912, 025.

\bibitem[\protect\citeauthoryear{Cruz Perez and Sol\`a}{2018}]{sola2}
Cruz P\'erez J. and Sol\`a J., 2018, Mod. Phys. Lett. A33, 1850228.

\bibitem[\protect\citeauthoryear{Damour}{2012}]{Damour}
Damour T., Class. Quant. Grav. 29 (2012) 184001.

\bibitem[\protect\citeauthoryear{Das and Banerjee}{2008}]{sudiptadi}
Das S. and Banerjee N., 2008, Phys. Rev. D 78, 043512.

\bibitem[\protect\citeauthoryear{Das and Banerjee}{2006}]{sudiptadigrg}
Das S. and Banerjee N., 2006, Gen. Relativ. Gravit., 38, 785.

\bibitem[\protect\citeauthoryear{Das et. al.}{2006}]{dascora}
Das S., Corasaniti P. and Khoury J., 2006, Phys. Rev. D, 73, 083509.

\bibitem[\protect\citeauthoryear{Delubac et. al.}{2015}]{delu}
Delubac T. et. al., 2015, Astron. Astrophys. 574, A59.

\bibitem[\protect\citeauthoryear{Dunajski and Gibbons}{2008}]{duna}
Dunajski M. and Gibbons G., 2008, Class. Quant. Grav. 25, 235012.

\bibitem[\protect\citeauthoryear{Elizalde et. al.}{2008}]{elizalde}
Elizalde E., Nojiri S., Odintsov S. D., S\'aez-G\'omez D. and Faraoni V., 2008, Phys. Rev. D. 77, 106005.

\bibitem[\protect\citeauthoryear{Evans et. al.}{2005}]{evans}
Evans A., Wehus I., Gr\o{}n, \O{} and Elgar\o{}y, \O{}., 2005, Astron. Astrophys. 430, 399.

\bibitem[\protect\citeauthoryear{Faraoni}{1999}]{faraoni}
Faraoni V., 1999, Phys. Rev. D 59, 084021.

\bibitem[\protect\citeauthoryear{Faraoni}{2005}]{fara}
Faraoni V., 2005, Ann. Phys. 317, 366.

\bibitem[\protect\citeauthoryear{Farooq and Ratra}{2013}]{farooq}
Farooq O. and Ratra B., 2013, Astrophys. J. 766, L7.

\bibitem[\protect\citeauthoryear{Farrar and Peebles}{2004}]{interdmde1}
Farrar G. R. and Peebles P. J. E., 2004, Astrophys. J. 604, 1.

\bibitem[\protect\citeauthoryear{Fierz}{1956}]{fierz}
Fierz M., Helv. Phys. Acta. 29 (1956) 128.

\bibitem[\protect\citeauthoryear{Foreman-Mackey et. al.}{2013}]{emcee}
Foreman-Mackey D., Hogg D. W., Lang D. and Goodman J., 2013, Publ. Astron. Soc. Pac. 125, 306.

\bibitem[\protect\citeauthoryear{Germani and Kehagias}{2010}]{germani}
Germani C. and Kehagias A., Phys. Rev. Lett. 2010, 105, 011302.

\bibitem[\protect\citeauthoryear{Gibbons and Hawking}{1977}]{gibbon}
Gibbons G. W. and Hawking S. W., 1977, Phys. Rev. D 15, 2738.

\bibitem[\protect\citeauthoryear{Gon{\c{c}}alves and Moss}{1997}]{goncamoss}
Gon{\c{c}}alves S. and Moss I., 1997, Class. Quant. Gravit. 14, 2607.

\bibitem[\protect\citeauthoryear{Guth}{1981}]{Gut81}
Guth A., 1981, Phys. Rev. D 23, 347.

\bibitem[\protect\citeauthoryear{Holden and Wands}{1998}]{holden}
Holden D. J. and Wands D., 1998, Class. Quant. Grav. 15, 3271.

\bibitem[\protect\citeauthoryear{Holsclaw et. al.}{2010}]{hols}
Holsclaw T., Alam U., Sans\'o B., Lee H., Heitmann K., Habib S. and Higdon D., 2010, Phys. Rev. Lett. 105,
241302 ; Phys. Rev. D 82, 103502.

\bibitem[\protect\citeauthoryear{Huey and Wandelt}{2006}]{interdmde2}
Huey G. and Wandelt B. D., 2006, Phys. Rev. D74, 023519.

\bibitem[\protect\citeauthoryear{Ishida and De Souza}{2011}]{ishida}
Ishida E. E. O. and De Souza R. S., 2011, Astron. Astrophys. 527, A49.

\bibitem[\protect\citeauthoryear{Jacobson}{1995}]{jacobson}
Jacobson T., 1995, Phys. Rev. Lett. 75, 1260.

\bibitem[\protect\citeauthoryear{Jamil et. al.}{2010}]{jamil}
Jamil M., Saridakis E. N. and Setare M. R., 2010, JCAP 1011, 032.

\bibitem[\protect\citeauthoryear{Kamenshchik et. al.}{2001}]{dmdeint1}
Kamenshchik A. Y., Moschella U. and Pasquier V., 2001, Phys. Lett. B511, 265.

\bibitem[\protect\citeauthoryear{Khoury and Weltman}{2004}]{khoury}
Khoury J. and Weltman A., 2004, Phys. Rev. Lett., 93, 171104 ; Phys. Rev. D, 69, 044026.

\bibitem[\protect\citeauthoryear{Kolitch and Eardley}{1995}]{koli}
Kolitch S. and Eardley D., 1995, Ann. Phys. (N.Y.) 241, 128.

\bibitem[\protect\citeauthoryear{La and Steinhardt}{1989}]{la}
La D. and Steinhardt P., 1989, Phys. Rev. Lett. 62, 376.

\bibitem[\protect\citeauthoryear{Masina and Notari}{2012}]{masina}
Masina I. and Notari A., Phys. Rev. Lett. 2012, 108, 191302.

\bibitem[\protect\citeauthoryear{Mathiazhagan and Johri}{1984}]{mathi}
Mathiazhagan C. and Johri  V. B., 1984, Class. Quant. Grav. 1, L29.

\bibitem[\protect\citeauthoryear{Micheletti et. al.}{2009}]{interdmde3}
Micheletti S., Abdalla E. and Wang B., 2009, Phys. Rev. D79, 123506.

\bibitem[\protect\citeauthoryear{Moresco et. al.}{2012}]{ohd4}
Moresco M., Verde L., Pozzetti L., Jimenez R. and Cimatti A., 2012, J. Cosmol. Astropart. Phys 07, 053.

\bibitem[\protect\citeauthoryear{Mota and Barrow}{2004}]{mota}
Mota D. and Barrow J., 2004, Mon. Not. R. Astron. Soc. 349, 291; Phys. Lett. B 581, 141.

\bibitem[\protect\citeauthoryear{Mota and Shaw}{2006}]{motashaw1}
Mota D. and Shaw D., 2006, Phys. Rev. Lett. 97, 151102.

\bibitem[\protect\citeauthoryear{Mota and Shaw}{2007}]{motashaw2}
Mota D. and Shaw D., 2007, Phys. Rev. D 75, 063501.

\bibitem[\protect\citeauthoryear{Mukherjee and Banerjee}{2017}]{mukherjee1}
Mukherjee A. and Banerjee N., 2017, Class. Quant. Grav. 34, 035016.

\bibitem[\protect\citeauthoryear{Mukherjee and Banerjee}{2016}]{mukherjee2}
Mukherjee A. and Banerjee N., 2016, Phys. Rev. D 93, 043002.

\bibitem[\protect\citeauthoryear{Mukherjee et. al.}{2019}]{mukherjee3}
Mukherjee A., Paul N. and Jassal H., 2019, JCAP 01, 005.

\bibitem[\protect\citeauthoryear{Mukherjee and Chakrabarti}{2019}]{purbasoumya}
Mukherjee P. and Chakrabarti S., Eur. Phys. J. C. 2019, 79 : 681.

\bibitem[\protect\citeauthoryear{Nordtvedt (Jr)}{1970}]{nordt}
Nordtvedt (Jr) K., 1970, Astrophys. J. 161, 1059.

\bibitem[\protect\citeauthoryear{Onofrio and Wegner}{2014}]{wegner2}
Onofrio R. and Wegner G. A., Astrophysical Journal, 2014, 791, 125.

\bibitem[\protect\citeauthoryear{Padmanabhan and Roychoudhury}{2003}]{paddy}
Padmanabhan T. and Roychoudhury T., 2003, Mon. Not. R. Astron. Soc., 344, 823.

\bibitem[\protect\citeauthoryear{Pan et. al.}{2018}]{pan}
Pan S., Mukherjee A. and Banerjee N., 2018, Mon. Not. Roy. Astron. Soc. 477, 1189.

\bibitem[\protect\citeauthoryear{Peebles and Ratra}{2003}]{peebles}
Peebles P. and Ratra B., 2003, Rev. Mod. Phys. 75, 559.

\bibitem[\protect\citeauthoryear{Perlmutter}{1999}]{Per99}
Perlmutter S. et al, 1999, ApJ. 517 (2): 565.

\bibitem[\protect\citeauthoryear{Riess}{1998}]{Rie98}
Riess A. et al, 1998, Astrophys. J. 116 (3): 1009.

\bibitem[\protect\citeauthoryear{Riess}{2001}]{riess}
Riess A., 2001, Astrophys. J., 560, 49.

\bibitem[\protect\citeauthoryear{Riess et. al.}{2004}]{riessobs}
Riess A. et. al., 2004, Astrophys. J. 607, 665.

\bibitem[\protect\citeauthoryear{Riess et. al.}{2018}]{riess2018}
Riess A. et al., 2018, ApJ. 855, 136.

\bibitem[\protect\citeauthoryear{Roy and Banerjee}{2017}]{nandan}
Roy N. and Banerjee N., 2017, Phys. Rev. D 95, 064048.

\bibitem[\protect\citeauthoryear{Ryan et. al.}{2018}]{ryan}
Ryan J., Doshi S. and Ratra B.,2018, Mon. Not. Roy. Astron. Soc. 480, no. 1, 759.

\bibitem[\protect\citeauthoryear{Sahni et. al.}{2003}]{vsahni}
Sahni V., Saini T., Starobinsky A. and Alam U., 2003, JETP Lett. 77 : 201 ; Pisma Zh. Eksp. Teor. Fiz 77 : 249.

\bibitem[\protect\citeauthoryear{S\'a}{2020}]{saa}
S\'a P., 2020, Phys. Rev. D. 102, 103519.

\bibitem[\protect\citeauthoryear{S\'a}{2020}]{saa1}
S\'a P., 2020, Universe, 6, 78.

\bibitem[\protect\citeauthoryear{Santos and Gregory}{1997}]{santos}
Santos C. and Gregory R., 1997, Ann. Phys. 258, 111.

\bibitem[\protect\citeauthoryear{Seikel, Clarkson and Smith}{2012}]{seikel}
Seikel M., Clarkson C. and Smith M., 2012, JCAP 1206, 036.

\bibitem[\protect\citeauthoryear{Shafieloo, Kim and Linder}{2012}]{shafi}
Shafieloo A., Kim A. and Linder E., 2012, Phys. Rev. D 85, 123530.

\bibitem[\protect\citeauthoryear{Simon, Verde and Jimenez}{2005}]{ohd1}
Simon J., Verde L. and Jimenez R., 2005, Phys. Rev. D. 71, 123001.

\bibitem[\protect\citeauthoryear{Sen and Sen}{2001}]{sensen}
Sen S. and Sen A. A., 2001, Mon. Not. R. Astron. Soc. 63, 124006.

\bibitem[\protect\citeauthoryear{Sol\`a et. al.}{2017}]{sola}
Sol\`a J., Karimkhani E. and Khodam-Mohammadi A., 2017, Class. Quant. Grav. 34, no.2, 025006.

\bibitem[\protect\citeauthoryear{Sol\`a}{2018}]{sola1}
Sol\`a J., 2018, Int. J. Mod. Phys. D. 27, no.14, 1847029.

\bibitem[\protect\citeauthoryear{Sol\`a et. al.}{2019}]{sola3}
Sol\`a J., G\'omez-Valent A., Cruz P\'erez J. and Moreno-Pulido C., 2019, Astrophys. J. 886, no.1, L6.

\bibitem[\protect\citeauthoryear{Sol\`a et. al.}{2020}]{sola4}
Sol\`a J., G\'omez-Valent A., Cruz P\'erez J. and Moreno-Pulido C., 2020, arXiv:2006.04273v3 [astro-ph.CO].

\bibitem[\protect\citeauthoryear{Sol\`a}{2015}]{solampla}
Sol\`a J., Mod. Phys. Lett. A. 2015, 30, 1502004.

\bibitem[\protect\citeauthoryear{Sol\`a}{2015}]{solag1}
Sol\`a J., Int. J. Mod. Phys. D. 2015, 24, 1544027.

\bibitem[\protect\citeauthoryear{Sol\`a and Gomez-Valent}{2015}]{solag2}
Sol\`a J. and G\'omez-Valent A., Int. J. Mod. Phys. D. 2015, 24, 1541003.

\bibitem[\protect\citeauthoryear{Sol\`a and Yu}{2020}]{solathermo}
Sol\`a J. and Yu H., 2020, Gen. Rel. Grav. 52, no.2, 17.

\bibitem[\protect\citeauthoryear{Stern et. al.}{2010}]{ohd2}
Stern D., Jimenez R., Verde L., Kamionkowski M. and Stanford S. A., 2010, J. Cosmol. Astropart. Phys 02, 008.

\bibitem[\protect\citeauthoryear{Tsujikawa et. al.}{2013}]{tsuji}
Tsujikawa S., Ohashi J., Kuroyanagi S. and De Felice A., Phys. Rev. D. 2013, 88, 023529 .

\bibitem[\protect\citeauthoryear{Van den Bergh}{1982}]{vdb}
Van den Bergh N., 1982, Gen. Relativ. Gravit. 14, 17.

\bibitem[\protect\citeauthoryear{Visser}{2005}]{visser}
Visser M., 2005, Gen. Rel. Grav. 37, 1541.

\bibitem[\protect\citeauthoryear{Wagoner}{1970}]{wagoner}
Wagoner R. V., 1970, Phys. Rev. D 1, 3209.

\bibitem[\protect\citeauthoryear{Wegner and Onofrio}{2015}]{wegner1}
Wegner G. A. and Onofrio R., Eur. Phys. J. C. 2015, 75, 307.

\bibitem[\protect\citeauthoryear{Weinberg}{1972}]{weinbook}
Weinberg S. 1972 \textit{Gravitation and Cosmology} (New York: Wiley).

\bibitem[\protect\citeauthoryear{Will}{1981}]{will}
Will C. M., 1981, Theory and Experiment in Gravitational Physics (Cambridge University, Cambridge, England).

\bibitem[\protect\citeauthoryear{Xianyu, Ren and He}{2013}]{xian}
Xianyu Z., Ren J. and He H-J., Phys. Rev. D. 2013, 88, 096013.

\bibitem[\protect\citeauthoryear{Zhang et. al.}{2014}]{ohd6}
Zhang C., Zhang H., Yuan S., Zhang T. J. and Sun Y. C., 2014, Res. Astron. Astrophys. 14, 1221.

\end{thebibliography}

\end{document}